\documentstyle[12pt]{article}
\def\hybrid{\topmargin -20pt	\oddsidemargin 0pt
	\headheight 0pt	\headsep 0pt
	\textwidth 6.25in	
	\textheight 9.5in	
	\marginparwidth .875in
	\parskip 5pt plus 1pt	\jot = 1.5ex}

\hybrid

\begin{document}
\def\x{\times}
\def\ra{\rightarrow}
\def\beq{\begin{equation}}
\def\eeq{\end{equation}}
\def\beqa{\begin{eqnarray}}
\def\eeqa{\end{eqnarray}}
\def\D{ {\cal D}}
\def\L{ {\cal L}}
\def\C{ {\cal C}}
\def\N{ {\cal N}}
\def\calE{{\cal E}}
\def\lin{{\rm lin}}
\def\Tr{{\rm Tr}}
\def\mxth{\mathsurround=0pt }
\def\xversim#1#2{\lower2.pt\vbox{\baselineskip0pt \lineskip-.5pt
x  \ialign{$\mxth#1\hfil##\hfil$\crcr#2\crcr\sim\crcr}}}
\def\simgr{\mathrel{\mathpalette\xversim >}}
\def\simle{\mathrel{\mathpalette\xversim <}}

\def\R{ {\cal R}}
\def\I{ {\cal I}}
\def\calR{ {\cal R}}
\def\lag{Lagrangian}
\def\Kahler{K\"{a}hler}

\def\a{\alpha}
\def\b{\beta}
\def\dota{ {\dot{\alpha}} }
\def\lag{Lagrangian}
\def\Kahler{K\"{a}hler}
\def\kahler{K\"{a}hler}
\def\A{ {\cal A}}
\def\C{ {\cal C}}
\def\D{ {\cal D}}
\def\F{{\cal F}}
\def\L{ {\cal L}}
\def\R{ {\cal R}}
\def\x{ \times }
\def\beq{\begin{equation}}
\def\eeq{\end{equation}}
\def\beqa{\begin{eqnarray}}
\def\eeqa{\end{eqnarray}}
\def\S4{\frac{SO(4,2)}{SO(4) \otimes SO(2)}}
\def\P3{\frac{SO(3,2)}{SO(3) \otimes SO(2)}}
\def\MGd{\frac{SO(r,p)}{SO(r) \otimes SO(p)}}
\def\SOd{\frac{SO(r,2)}{SO(r) \otimes SO(2)}}
\def\SO2{\frac{SO(2,2)}{SO(2) \otimes SO(2)}}
\def\SUm{\frac{SU(n,m)}{SU(n) \otimes SU(m) \otimes U(1)}}
\def\SUS{\frac{SU(n,1)}{SU(n) \otimes U(1)}}
\def\SK{\frac{SU(2,1)}{SU(2) \otimes U(1)}}
\def\SU{\frac{ SU(1,1)}{U(1)}}

\renewcommand{\thesection}{\arabic{section}.}
\renewcommand{\theequation}{\thesection \arabic{equation}}

\begin{titlepage}
\begin{center}

\hfill HUB-IEP-94/20\\
\hfill UPR-634T\\
\hfill hep-th/9410056\\

\vskip .1in

{\bf MODULI DEPENDENT
NON-HOLOMORPHIC CONTRIBUTIONS OF MASSIVE STATES TO
GRAVITATIONAL COUPLINGS AND \\
$C^2$-TERMS IN $Z_N$-ORBIFOLD
COMPACTIFICATIONS}

\vskip .2in

{\bf Gabriel Lopes Cardoso,
Dieter L\"{u}st }  \\
\vskip .1in

{\em  Humboldt Universit\"at zu Berlin\\
Institut f\"ur Physik\\
D-10115 Berlin, Germany}\footnote{e-mail addresses:
GCARDOSO@QFT2.PHYSIK.HU-BERLIN.DE, LUEST@QFT1.PHYSIK.HU-BERLIN.DE}

\vskip .1in
and
\vskip .1in

{\bf Burt A. Ovrut}
\vskip .1in

{\em
Department of Physics\\
University of Pennsylvania\\
Philadelphia, PA 19104-6396,
USA}\footnote{e-mail address: OVRUT@OVRUT.HEP.UPENN.EDU}

\end{center}

\vskip .2in

\begin{center} {\bf ABSTRACT } \end{center}
\begin{quotation}\noindent

It is pointed out that massive states in D=4, N=1 supergravity-matter
theories
can, in general, at the 1-loop level contribute non-holomorphic
terms to quadratic gravitational couplings.  It is then shown
in the context of $(2,2)$-symmetric $Z_N$-orbifold theories
that,
for constant moduli backgrounds,
the inclusion of such contributions can result in the cancellation
of naked $C^2$-terms.  ${\cal R}^2$-terms can also arise but, being
ghost free, need not cancel.

\end{quotation}
October 1994\\
\end{titlepage}
\vfill
\eject

\newpage

\section{Introduction}

\hspace*{.3in}
Threshold corrections to gauge couplings
have been extensively studied in the past
 in the context of $(2,2)$-symmetric  $Z_N$-orbifold
theories
\cite{Vkap,Dix,10,12,DFKZ,9,40,May,May2,Bailin1,Bailin2}.  These
threshold corrections, which arise
when integrating out the massive modes of the string, are moduli
dependent, since the values of
some of these masses depend on the moduli.
Moduli dependent threshold corrections
play a crucial role in string unification as well as for the problem
of supersymmetry breaking.  Similarly,
threshold corrections
to gravitational couplings also have been addressed
in the past \cite{13,40,15}.
The gravitational case is, however, more complicated
than the gauge case in that there are in general
three different quadratic curvature
terms.  Thus, one has to study 1-loop corrections to
three different quadratic gravitational coupling functions.
Any quadratic curvature term
can be expressed as a linear combination of ${\cal R}^2$, ${\cal R}^{mn}
{\cal R}_{mn}$ and $C^2$ terms.  $C^2$ denotes the square of the Weyl
tensor.  From a supergravity point of view, this set forms a natural basis
of quadratic curvature terms, since
${\cal R}^2$, ${\cal R}^{mn}
{\cal R}_{mn}$ and $C^2$ are contained in the highest component of
the supergravity
superfields $\bar{R} R$, $G^2_{\alpha \dot{\alpha}}$ and
$W^2_{\alpha \beta \gamma}$, respectively.
When discussing the physics of quadratic curvature terms, however,
it is convenient to revert to a different basis, namely to the one
spanned by ${\cal R}^2$, $C^2$ and the Gauss-Bonnet combination $GB$.
It is known \cite{Stelle,FERN} that the following Lagrangian,
${\cal L}_{{\cal R}^2} =
-\frac{1}{2 {\kappa}^2} {\cal R} + \alpha {\cal R}^2$, describes the coupling
of a physical scalar mode of mass
$m^2 \sim \frac{1}{\alpha {\kappa}^2}$ to Einstein
gravity.  On the other hand, it is also known \cite{Stelle,FERN} that the
Lagrangian
${\cal L}_{C^2}= -\frac{1}{2 {\kappa}^2} {\cal R} + \beta C^2$ describes
the coupling of ghost modes with mass
$m^2 \sim \frac{1}{\beta {\kappa}^2}$ to
Einstein gravity.  Thus, Lagrangian ${\cal L}_{C^2}$ is usually
viewed as describing undesirable physics.  In fact, it has
been
argued in \cite{ZWIEB} that
the appearance of such ghost modes in string theory would violate unitarity
and, hence, naked $C^2$-terms shouldn't occur in effective string theory
Lagrangians.  It was pointed out in \cite{GROSS,TSEYT}, however, that since
these ghost modes are at the Planck mass, they
really are in a region of large momentum, for which the perturbative
$\alpha'$ expansion of the effective orbifold Lagrangian is unreliable.
Thus, in contrast to field theory
there is a priori no reason for forbidding the appearance
of naked $C^2$-terms in the effective Lagrangian of superstring theory.

Nevertheless, the conventional choice \cite{GROSS,TSEYT}
of tree-level couplings in
string theory to quadratic gravitational curvature terms
is taken to be the one which describes the coupling of the dilaton
to the Gauss-Bonnet combination, only.
This conventional choice, to which we will
stick throughout this paper,
can be looked upon as a gauge choice \cite{TSEYT}, because
additional couplings of the dilaton to $C^2$ and to ${\cal R}^2$
can be removed by an appropriate field redefinition of the dilaton
multiplet.
In the S-matrix approach this ambiguity is a consequence of an
off-shell ambiguity in the subtraction of the exchanges involving
massless modes.
Thus, the conventional choice
assumes that there is a particular subtraction scheme which, in the tree-
level Lagrangian, translates into a coupling of the
dilaton multiplet to the Gauss-Bonnet combination, only.

Now, let us consider 1-loop
moduli dependent threshold corrections to gravitational couplings
in $Z_N$-orbifold theories.  As discussed above, there is a priori
nothing wrong with generating threshold corrections proportional to $C^2$.
As we will show, the following must hold.
The moduli dependent threshold corrections to
couplings are due to both massless
and massive particles running in the loop.  The massless contributions
can be calculated in field theory using the manifestly supersymmetric
procedure introduced in \cite{15}.  Such contributions are non-local and
non-holomorphic
and found to be proportional not only
to ${\cal R}^2$ and to $GB$, but also to $C^2$.  When inserting vev's
for the
moduli background fields, these contributions turn into
local non-holomorphic terms.
We will also argue that
there are two types of contributions due to the massive modes.  The first
type consists of
local holomorphic
contributions proportional to the chiral masses of (some restricted set
of) massive states.  The second type consists of
local non-holomorphic contributions
to the gravitational couplings of
${\cal R}^2$ and ${\cal R}^{mn} {\cal R}_{mn}$.
An analysis based on dimensional grounds shows that,
in fact, it is a priori
perfectly possible
to have local non-holomorphic gravitational couplings to ${\cal R}^2$ and
to ${\cal R}^{mn}{\cal R}_{mn}$ compatible with $N=1$ supersymmetry.
Then, rewriting ${\cal R}^{mn} {\cal R}_{mn}$ into $C^2$ and $GB$-terms
yields massive non-holomorphic
contributions to $C^2$.  Note that the massive particles never produce
non-local contributions to the gravitational couplings.  This is in contrast
to the massless contributions which are explicitly non-local for non-constant
background moduli.
Thus, the effective Lagrangian
will in general
contain both non-local and local terms proportional to
$C^2$.

The tree-level Lagrangian of
$Z_N$-orbifold theories is invariant under duality transformations.  Threshold
corrections to gravitational couplings are expected to respect
duality invariance.  Duality invariance of the thresholds can be achieved
by taking into account the massive contributions of the first type, namely
the local holomorphic contributions proportional to the chiral masses of
some set of massive states.

As argued above, the effective Lagrangian of $(2,2)$-symmetric
$Z_N$-orbifold theories will in general
contain both non-local and local terms
proportional to $C^2$, steming from the massless and the massive
contributions, respectively.  For constant moduli background fields, the
non-local contributions turn into local ones, and the resulting effective
Lagrangian will then only contain local $C^2$-terms with a constant
moduli dependent coefficient.  As stated above, there is a priori
nothing wrong with such a naked $C^2$-term in the effective orbifold
Lagrangian.  Thus, its coefficient need a priori not be zero.  Let us
now nevertheless
consider the possibility that, in an actual string calculation, this
coefficient is actually found to be vanishing.  Then,
for this to be the case, it is crucial that
 the massive states contribute local
non-holomorphic
terms
of the type discussed above. Also note that even if there is such
an exact cancellation of naked $C^2$-terms for constant moduli background
fields, this cancellation doesn't hold anymore for non-constant
moduli background fields, and the effective Lagrangian will contain
$C^2$-terms with both non-local and local non-holomorphic moduli
dependent functions.

The paper is organised as follows.  In section 2
the most general local K\"{a}hler invariant superfield
Lagrangian with
quadratic curvature terms
is given.  It is followed by a discussion of the properties of
the
associated
gravitational coupling functions.
Section 3 contains a discussion of the tree-level Lagrangian
of $(2,2)$ symmetric orbifold theories with emphasis on the tree-level
couplings of the dilaton to quadratic curvature terms.  It is followed
by a general discussion of 1-loop corrections to the gravitational couplings
compatible with $N=1$ supersymmetry.

In section 4
the
manifestly supersymmetric procedure introduced in \cite{15}
for calculating mixed gravitational-
K\"{a}hler and mixed gravitational-$\sigma$ model anomalies
in field theory is reviewed.  It consists of first performing
the calculations in conventional superspace and then, at the end,
reverting to K\"{a}hler superspace.  In conventional superspace the
relevant symmetry is the super-Weyl-K\"{a}hler symmetry which, at
the 1-loop level, gets broken by anomalies.  The relevant graphs are
the ones for computing mixed gravitational-super-Weyl-K\"{a}hler
as well as mixed gravitational-$\sigma$-model anomalies.  Then,
rotating the resulting superfield expressions over to
K\"{a}hler superspace yields the mixed
gravitational-
K\"{a}hler and mixed gravitational-$\sigma$ model anomalies.
We apply this procedure to
$Z_N$-orbifolds in order to calculate
the 1-loop moduli dependent contributions
to gravitational couplings due to the massless modes
running in the loop.  We introduce a suitable parametrisation of
those massless 1-loop contributions which we haven't computed explicitly,
such as the one from the supergravity multiplet.
Sticking to the conventional choice \cite{GROSS,TSEYT} for the tree-level
gravitational coupling of the dilaton, we then show, in
Appendix B, that
the massless modes will, in general, contribute
to a naked $C^2$-term; that is to a $C^2$-term which is not contained
in the Gauss-Bonnet combination.

In section 5,
a suitable
parametrisation of the moduli dependent 1-loop contributions to the
gravitational couplings from the massive modes
is introduced.  We argue that such
contributions may not only
occur as local holomorphic
contributions proportional to the chiral masses of (some restricted set
of) massive states, but also
as local non-holomorphic contributions
to the gravitational couplings of
${\cal R}^2$ and ${\cal R}^{mn} {\cal R}_{mn}$.
Upon rewriting
${\cal R}^{mn} {\cal R}_{mn}$ into $C^2$ and $GB$-terms, we show
that
massive modes may contribute local non-holomorphic terms to naked
$C^2$-terms.
Section 6 contains our conclusions.  And finally,
in Appendix A it is shown for the case of the $Z_4$-orbifold
that the field theoretical
calculation of one of the gravitational couplings
agrees with
the string scattering amplitude calculation \cite{13}
of the same gravitational coupling.

\section{$U_K(1)$ Superspace, Fourth-Order Superfield
Lagrangians and their Symmetries}

\hspace*{.3in} In this section, we will briefly review
some of the features of K\"{a}hler superspace geometry
which will
be relevant in the subsequent discussion.  We will also discuss the
structure of
the most general local fourth-order supergravity Lagrangian compatible with
the symmetries of K\"{a}hler superspace.

We begin with a short review of $U_K(1)$ superspace.
A complete description
of its properties can be found in
\cite{6}.  The structure group of \Kahler\ superspace is
taken to be $SL(2,C) \x U_K (1)$ and, accordingly, one
introduces two Lie algebra valued one-form gauge connections
$\phi_B \,^A = dz^M \phi_{MB}\,^A$ and $A = dz^M$ $A_M$
corresponding to the Lorentz and $U_K (1)$ groups, respectively.
In addition, one introduces a supervielbein $E_M \,^A$ and the
associated one-forms $E^A = dz^M E_M \,^A$.  The $U_K (1)$ gauge
connection $A$ is a composite gauge connection defined by
\beqa A_\alpha &=& \frac{1}{4} \D_\alpha K \nonumber \\
 A^{\dot{\alpha}} &=& -\frac{1}{4} \bar{\D}^{\dot{\alpha}} K
\nonumber \\
A_{\alpha \dot{\alpha}} &=& - \frac{i}{8} \left[ \D_{\alpha},
\bar{\D}_{\dot{\alpha}} \right] K   \eeqa   
where the prepotential $K(\Phi_i, \; \bar{\Phi}_{\bar{i}})$
is the \Kahler\
potential for matter chiral superfields $\Phi_i$.  All matter
superfields have vanishing $U_K (1)$ weight, $\omega_K(\Phi_i) =
0$.  Under a \Kahler\ transformation
\beq {\kappa}^2 K(\Phi_i,\; \bar{\Phi}_{\bar{i}})
\ra {\kappa}^2 K (\Phi_i, \; \bar{\Phi}_{\bar{i}}) +
F(\Phi_i) + \bar{F} (\bar{\Phi}_{\bar{i}})
\label{ktr}
\eeq 
the one-form $A$ transforms as
\beq A \ra A + {\kappa}^{-2} \frac{i}{2} d  \; Im \, F \eeq 
where $Im \, F = \frac{F- \bar{F}}{2i}$ and
${\kappa}^2 = 8 \pi M_P^{-2}$.  $M_P$ is the Planck mass.
Also, under a \Kahler\
transformation the supervielbein one-forms $E^A$ can be shown
\cite{6} to transform as
\beq E^A \ra E^A \exp \left[ - \frac{i}{2} \omega (E^A) Im\, F
\right] \eeq 
where
\beq \omega (E^\alpha) = 1, \; \; \omega (E_{\dot{\alpha}}) = -1
\;\; \omega (E^a) = 0\;\; . \eeq 
Solving the Bianchi identities subject to a set of constraints
\cite{60}, one finds that all components of the torsion and
curvature may be expressed in terms of a set of superfields and
their coordinate derivatives
\beq \begin{array}{lcccccc}
{\rm superfield} & \;\;\; & R & \bar{R} & G_{\alpha \dot{\alpha}} &
W_{\alpha \beta \gamma} \; , \; X_\alpha & \bar{W}_{\dot{\alpha}
\dot{\beta} \dot{\gamma}}\; , \; \bar{X}_{\dot{\alpha}} \\
U_K(1) \; {\rm weight} & & 2 & -2 & 0 & 1 & -1 \end{array}
\label{ukw}
\eeq
where
\beqa X_\alpha &=& \D_\alpha R - \bar{\D}^{\dot{\alpha}}
G_{\alpha \dot{\alpha}} = - \frac{\kappa^2}{8}\left( \bar{\D}^2 -
8R \right) \D_\alpha K \nonumber \\
\bar{X}^{\dot{\alpha}} &=& \bar{\D}^{\dot{\alpha}} \bar{R} +
\D_\alpha G^{\alpha \dot{\alpha}} = - \frac{\kappa^2}{8}
\left(\D^2 - 8\bar{R}\right) \bar{\D}^{\dot{\alpha}} K \eeqa 
$X_{\alpha}$ is the superfield fieldstrength of the $U_K(1)$
gauge connection.
Note that the three superfields $R, W_{\alpha \beta \gamma}$ and
$X_{\alpha}$ are chiral, that is
\beqa \bar{\cal{D}}_{\dot{\alpha}} R = 0 \;\;,\;
\bar{\cal{D}}_{\dot{\alpha}} W_{\alpha \beta \gamma} =0
\;\;,\;
\bar{\cal{D}}_{\dot{\alpha}} X_{\alpha}=0 \eeqa
If we further assume that there is an internal
gauge group, then we must introduce yet another Lie algebra
valued one-form gauge connection $\A_a\,^b = dz^M \A_{M
a}\,^b$.  Solving the Bianchi identities now introduces a new chiral
superfield fieldstrength, $W^a_\alpha$, with $U_K(1)$ weight
$\omega (W_\alpha^a) = 1$.

Using these superfields, one can write down the most
general Lorentz and gauge invariant
local tree-level
superfield \lag\ in K\"{a}hler superspace
\cite{4,5,6} as follows.  Its component expansion will be organised
in powers of spacetime derivatives and
we will be interested in all terms with up to
four spacetime derivatives in the graviton field
.  At the level of two spacetime derivatives
the tree-level
Lagrangian  consists of three
parts, each specified by a fundamental and independent function.
The first part, specified entirely by the K\"{a}hler potential, is
the supergravity-matter kinetic energy term given by
\beq \L_0 = -\frac{3}{2} \kappa^{-2} \int d^4 \theta E \left[K\right]
+ h.c.
\label{lzero}
\eeq
where $E$ is the superdeterminant.  The second part, specified by
the holomorphic superpotential $W(\Phi_i)$, is the potential
energy term given by
\beq \L_{PE} = \frac{1}{2} \int d^4 \theta \frac{E}{R}
e^{{\kappa}^2 \frac{K}{2}} W(\Phi_i) + h.c.
\label{pot}
\eeq 
Finally, the Yang-Mills \lag\ is given by
\beq \L_{YM} = \frac{1}{8} \int d^4 \theta \frac{E}{R}
f(\Phi_i)_{ab} \; W^{\alpha a} W_\alpha\,^b + h.c.
\label{lym}
\eeq 
where $f(\Phi_i)_{ab}$ is the holomorphic gauge coupling
function.
These three parts, (\ref{lzero}),
(\ref{pot}) and (\ref{lym}), are manifestly Lorentz
invariant.  They are also gauge invariant, as the K\"{a}hler potential $K$
is invariant under Yang-Mills transformations of the charged matter superfields
$\Phi_i$.

Next, let us write down the most general local Lorentz invariant Lagrangian of
fourth-order supergravity, with matter coupled to it.  It is given
by quadratic combinations of the superfields appearing in
(\ref{ukw})
and reads \cite{25,GRIMM1}
\beqa \L_{(fourth)} &=& \int d^4 \theta \frac{E}{R}
g(\Phi_i)\; W^{\alpha \beta \gamma}
W_{\alpha \beta \gamma} + \int d^4 \theta E
\Delta(\Phi_i, {\bar \Phi_i})\; G^{\alpha \dot{\alpha}}
G_{\alpha \dot{\alpha}}  \nonumber\\
&+& \int d^4 \theta E
\Sigma(\Phi_i, {\bar \Phi_i}) \; {\bar R} R +
\int d^4 \theta \frac{E}{R}
h(\Phi_i)\; X^{\alpha}
X_{\alpha}  +  h.c.
\label{fourtho}
\eeqa 
Note that the matter field dependent
functions multiplying the chiral superfields $
W^{\alpha \beta \gamma} W_{\alpha \beta \gamma}$ and
$X^{\alpha} X_{\alpha}$,
$g(\Phi_i)$ and $h(\Phi_i)$ respectively, are holomorphic.
This is to be contrasted with
the
functions multiplying the real superfields $G^{\alpha \dot{\alpha}}
G_{\alpha \dot{\alpha}} $ and ${\bar R} R$
which, in general, are
non holomorphic
functions of the matter fields $\Phi_i$.
Gauge invariance of (\ref{fourtho})
requires the functions $g, \Delta, \Sigma$ and $h$
to be gauge invariant.
Also note that the Lagrangian (\ref{fourtho}) has mass dimension four if
the matter field dependent functions have mass dimension zero.

Having thus written out the most general Lorentz and gauge invariant tree level
Lagrangian with terms up to four
spacetime derivatives, we proceed in discussing
its transformation properties under K\"{a}hler transformations
(\ref{ktr}).
The tree level terms (\ref{lzero}),
(\ref{pot}) and (\ref{lym}) are invariant under
K\"{a}hler transformations (\ref{ktr})
by virtue of the transformations laws
\cite{4,5,6}
\beqa E & \ra & E \nonumber \\
R & \ra & e^{-(F-\bar{F})/2} R \nonumber\\
W & \ra & e^{-F} W \nonumber \\
W_\alpha & \ra & e^{-(F-\bar{F})/4} W_\alpha
\label{mud}
\eeqa 
It follows from (\ref{ukw}) that the two chiral superfields $W_{\alpha
\beta \gamma}$ and $X_{\alpha}$ carry the same $U(1)_K$ weight as
$W_{\alpha}$, $ \omega_K = 1$.  Hence, under K\"{a}hler transformations
$W_{\alpha \beta \gamma}$ and $X_{\alpha}$
transform
in the same way as
$W_{\alpha}$ given in (\ref{mud}).
The superfields $G_{\alpha \dot{\alpha}}$ and $\bar{R} R$ are inert
under K\"{a}hler transformations, since both have vanishing K\"{a}hler
weight $\omega_K$.
Thus, the fourth-order supergravity Lagrangian (\ref{fourtho})
is K\"{a}hler invariant provided that the matter field dependent functions
$g, \Delta, \Sigma $ and $h$ in (\ref{fourtho}) are K\"{a}hler invariant.

It is useful to display some of the component level terms contained in
the K\"{a}hler invariant tree-level Lagrangian (\ref{lzero})-(\ref{lym}).
Component fields are defined
according to standard notation \cite{4,5,6}:  $A^i$,
$\chi^i_\alpha$, $\F^i$ for chiral multiplets (and similar
notations for antichiral multiplets) and $\lambda^\alpha$, $v_m$,
$\D$ for Yang-Mills multiplets.  The irreducible minimal
supergravity multiplet is realized by $(e_m\,^a, \;
\psi^\alpha_m, \; M, \; b_a)$.  $M$ and $b_a$ denote the
auxiliary component fields of minimal supergravity.

All component Weyl fermions transform under \Kahler\
transformations (\ref{ktr}).  The component matter Weyl fermion,
$\chi_\alpha^i = (\frac{1}{\sqrt{2}}) \D_\alpha \Phi_i|$,
transforms as
\beq \chi'^i = e^{(i/2 \; Im \, F)} \chi^i \eeq 
whereas the gaugino $\lambda^\alpha$ transforms with opposite
charge
\beq \lambda'^\alpha = e^{(- i/2 \; Im \, F)} \lambda^\alpha \eeq
as does the gravitino $\psi^\alpha_m$.
The component connection for gauging \Kahler\
transformations is given by the lowest component
\cite{4,5,6} of the $U_K(1)$ gauge connection superfield
${A}_{\alpha\dot{\alpha}}$
\beq \left. A_{\alpha \dot{\alpha}} \right| = - \frac{i}{8}
\left[ \D_\alpha, \bar{\D}_{\dot{\alpha}} \right] \left. K
\right| = a_{\alpha \dota} = \sigma^m_{\alpha \dota} a_m
\label{acon}
\eeq
where
\beq a_m = \frac{1}{4} \left( \partial_j K \D_m A^j -
\partial_{\bar{j}} K \D_m \bar{A}^{\bar{j}} \right) + i
\frac{1}{4} g_{i \bar{j}} \left(\chi^i \sigma^m
\bar{\chi}^{\bar{j}} \right) \eeq 
Here, $g_{i\bar{j}}$ denotes the \Kahler\ metric $g_{i \bar{j}}
= \partial_i \partial_{\bar{j}} K$ of the matter manifold
parameterized by $A^i$ and $\bar{A}^{\bar{j}}$.  Under \Kahler\
transformations (\ref{ktr})
\beq a_m\, ' = a_m + {\kappa}^{-2}
\frac{i}{2} \partial_m \; Im \, F.\eeq 

The covariant derivative for matter fermions
$\chi^i_\alpha$ reads \cite{4,5,6}
\beqa \D_m \chi^i & = & \left( \partial_m \right.
+ iv_m^{(r)} \left( T^{(r)} - \frac{1}{2}
\kappa^2 \frac{\partial K}{\partial A^{ia} }
T^{(r)a} \,_b \, A^{ib}
\right)
\nonumber \\ &-& \left. \omega_m + \frac{i}{2} b_m - \kappa^2
a_m \right) \chi^i
+ \Gamma^i_{jk} \D_m A^j \chi^k \eeqa 
Here,
$\Gamma^i_{jk} = g^{i \bar{j}}
\partial_j g_{k \bar{j}}$ denotes the
$\sigma$-model
Christoffel connection of the matter manifold parametrized by
$A^i$ and $\bar{A}^j$.  $\omega_m$ and $v^{(r)}_m$ denote the
component Lorentz and the component Yang-Mills connection,
respectively.  Note that only fermions rotate under K\"{a}hler
transformations (\ref{ktr}). Thus, only fermions couple to
the K\"{a}hler connection $a_m$ given in (\ref{acon}).
For instance, since the
K\"{a}hler charge $\omega_K (\Phi_i)$ of matter superfield $\Phi_i$
is zero, $\omega_K (\Phi_i) = 0$, the scalar field $A^i$ is inert
under K\"{a}hler transformations and, indeed, there is no coupling to
$a_m$ in the
covariant derivative of the matter scalar $A^i$
\beq \D_m A^i = \partial_m A^i + iv^{(r)}_m T^{(r)} A^i \eeq

The component expansion of the kinetic Lagrangian (\ref{lzero})
reads
\beqa \L_0/e &=& - \frac{1}{2} \kappa^{-2} e \R - \frac{1}{3}
\kappa^{-2} \left( M \bar{M} - b^a b_a \right) \nonumber \\
&-& g^{mn} g_{i \bar{j}} \D_n A^i \D_m \bar{A}^{\bar{j}} -
\frac{i}{2} \chi^{\alpha i} g_{i\bar{j}} \sigma^m_{\alpha \dota}
\D_m \bar{\chi}^{\bar{j} \dota} + \frac{i}{2} \left( \D_m
\chi^{\alpha i} \right) g_{i \bar{j}} \sigma^m_{\alpha \dota}
\bar{\chi}^{\bar{j} \dota} \nonumber \\
&+&  \ldots \eeqa
where we have only displayed the component terms relevant for
this paper.
Note that it immediately
gives the correctly normalized Einstein-Hilbert
action as well as making the component \Kahler\ structure in the
matter sector manifest.  This is, in fact,
one of the advantages of the \Kahler\ superspace formulation,
since
it immediately gives the correctly normalized
kinetic terms for all the component fields without any need for
rescalings or complicated partial integrations at the component
field level.

Next, we would like to display some of the component terms in the four
spacetime derivative Lagrangian (\ref{fourtho}).  The highest component
of $W^{\alpha \beta \gamma} W_{\alpha \beta \gamma}$ contains \cite{25}
\beqa
\int d^4 \theta \frac{E}{R}
W^{\alpha \beta \gamma}
W_{\alpha \beta \gamma}  = \frac{1}{8} \left( C^{mnpq} C_{mnpq}
- i \R_{mna}\,^b \; \tilde{\R}^{mn}\,_b\,^a \right) + \cdots
\eeqa
where $\tilde{\R}^{mn}\,_a\,^b = \frac{1}{2} \epsilon^{mnlp}
{\R}_{lpa}\,^b$ and
where $C^{mnpq} C_{mnpq}$ denotes
the square of the Weyl tensor, which , in four dimensions, is expressable
as
\beqa
C^{mnpq} C_{mnpq} = {\cal R}^{mnpq} {\cal R}_{mnpq} -
2 {\cal R}^{mn} {\cal R}_{mn} + \frac{1}{3} {\cal R}^2
\eeqa
The highest component of $G^{\alpha \dot{\alpha}} G_{\alpha \dot{\alpha}}$
contains \cite{25}
\beqa
\int d^4 \theta E
G^{\alpha \dot{\alpha}}
G_{\alpha \dot{\alpha}}  + h.c. =  \frac{1}{2} ({\cal R}^{mn}
{\cal R}_{mn} - \frac{2}{9} {\cal R}^2)
+ \cdots \eeqa
Finally, the highest component of $\bar{R} R$ contains \cite{25}
\beqa
\int d^4 \theta E
{\bar R} R + h.c = \frac{1}{72} {\cal R}^2 + \cdots   \eeqa
We have not displayed the highest component of $X^{\alpha} X_{\alpha}$,
as we will not consider such couplings in this paper.

Note that the ${\tilde {\cal R}}{\cal R}$ term is contained in
the highest component of $W^{\alpha \beta \gamma}W_{\alpha \beta \gamma}$
,only.
Also note that, from
a supergravity point of view, $C^2$,
${\cal R}^{mn}{\cal R}_{mn}$ and ${\cal R}^2$
form a natural basis of the vector space of quadratic curvature terms.
The component expansion of the four spacetime
derivative Lagrangian (\ref{fourtho})
then reads
\beqa
{\cal L} &=&  \frac{1}{4} {\it Re}\, g(A^i) C^{mnpq} C_{mnpq}
+ \frac{1}{2}
\Delta(A^i, \bar{A^i}) ({\cal R}^{mn} {\cal R}_{mn} - \frac{2}{9}
{\cal R}^2)
+ \frac{1}{72}
\Sigma(A^i, {\bar A}^i) {\cal R}^2 \nonumber\\
&+& \frac{1}{4} {\it Im}\, g(A^i)  \R_{mna}\,^b \; \tilde{\R}^{mn}\,_b\,^a  +
\cdots \eeqa
where we have only displayed the component terms relevant for this
paper.

Instead of using $C^2$, ${\cal R}^{mn} {\cal R}_{mn}$ and ${\cal R}^2$
as a set of basis vectors spanning the vector space of quadratic
curvature terms, one can also use another set of linearly independent
vectors given by $C^2$, $GB$ and ${\cal R}^2$.
$GB$ denotes
the Gauss-Bonnet combination
\beqa
GB = C^{mnpq} C_{mnpq} - 2
{\cal R}^{mn} {\cal R}_{mn} + \frac{2}{3} {\cal R}^2 \eeqa
and is
contained in the highest component of the following superfield
\beqa
GB =
\left( 8 W^{\alpha \beta \gamma} W_{\alpha \beta \gamma}
+ (\bar{{\cal D}}^2 - 8 R) (G^{\alpha \dot{\alpha}} G_{\alpha \dot{\alpha}}
- 4 \bar{R} R)
\right)|_{\theta^2}
+ h.c
\label{sgb}
\eeqa
Both sets of base vectors will play an important role in the subsequent
discussion of 1-loop corrections to gravitational couplings.

\section{Effective gravitational couplings in (2,2) symmetric orbifold
theories}

\hspace*{.3in} In this section, we will consider (2,2) symmetric
$Z_N$ orbifolds \cite{DHVW1,DHVW2,FerPor}.
We will, for simplicity,
restrict our discussion to those
orbifolds which do not contain (1,2) moduli.
Our considerations can, however, be
generalized in a straightforward way to the case of the remaining
orbifolds, which do contain (1,2) moduli.

Generically, the massless
spectrum
\cite{FerPor} of the orbifolds under consideration contains
a set of
uncharged untwisted (1,1) moduli fields, $T^{IJ}$,
describing the geometry of the underlying six-torus.
Among them, the
three diagonal ones
are denoted by $T^{II}=T^I$.
The massless
spectrum
also contains
twisted (1,1) moduli as well as untwisted (charged
or uncharged) and twisted (charged or uncharged)
matter fields. The untwisted off-diagonal moduli $T^{IJ}, I \neq J,$
can be regarded as additional matter fields, and we will do so
in the following.
All of these fields, other than the $T^I$, will
be collectively denoted
by $\phi^i$.
In addition to the above fields, there is also
the universal dilaton supermultiplet, present in any compactification
scheme of the heterotic superstring theory.

Recall that the most general tree-level K\"{a}hler invariant
supergravity-matter Lagrangian containing quadratic curvature terms
is given by
\beqa
\L &=& - \frac{3}{2} \kappa^{-2} \int d^4 \theta E \left[K\right]
+ \frac{1}{2} \int d^4 \theta \frac{E}{R}
e^{{\kappa}^2 \frac{K}{2}} W(\Phi) \nonumber\\
&+& \frac{1}{8} \int d^4 \theta \frac{E}{R}
f(\Phi)_{ab} \; W^{\alpha a} W_\alpha\,^b
+ \int d^4 \theta \frac{E}{R}
g(\Phi)\; W^{\alpha \beta \gamma}
W_{\alpha \beta \gamma} \nonumber\\
&+& \int d^4 \theta E
\Delta(\Phi, {\bar \Phi})\; G^{\alpha \dot{\alpha}}
G_{\alpha \dot{\alpha}}
+ \int d^4 \theta E
\Sigma(\Phi, {\bar \Phi}) \; {\bar R} R +  h.c.
\label{orblag}
\eeqa 
In the following, we will not be concerned with the superpotential $W$
and so we will omit it.
What do the functions $K,f,g,\Delta$ and $\Sigma$ look like
for the (2,2) symmetric orbifold tree-level
Lagrangians under consideration?  The
K\"{a}hler potential
$K$ can be generically expanded in powers of $\phi^i$ as
\beqa K(T,\bar{T},S,\bar{S},\phi,\bar{\phi}) =
K_0 +{\cal Z}_{ij} (T, \bar{T})
\bar{\phi}^i \phi^j  + {\cal O}
\left( \left( \bar{\phi} \phi \right)^2 \right)
\label{kpot}
\eeqa
where
\beqa   \kappa^{2} K_0 &=& - ln (S + \bar{S}) -
 \sum_I ln (T + \bar{T})^I
\label{kts}
\eeqa
$S$ denotes the chiral representation of the
universal dilaton supermultiplet.
Also,
${\kappa}^2 = 8\pi M^{-2}_P$ , where $M_P$ is the Planck mass.
Note that in our conventions the
chiral matter superfields $\phi^i$ have canonical dimension one, while
the dilaton $S$  and the untwisted moduli $T^I$ are dimensionless.
The metric ${\cal Z}_{ij}$, which does not depend on the dilaton $S$, has
the following moduli dependence
\beqa           {\cal Z}_{ij} \left(T^I, \bar{T}^I\right) &=&
\delta_{ij} \prod_I
\left[ T^I + \bar{T}^I \right]^{ q_I^i} \eeqa
where the exponents $q_I^i$ depend on the
particular matter $\phi^i$.

At tree-level, the holomorphic gauge coupling function $f_{ab}$ is given
by
\beqa
f_{ab} = {\delta}_{ab} k_a S
\label{fab}
\eeqa
where $k_a$ denotes the level of the Kac-Moody gauge algebra.
The holomorphic gravitational coupling function $g$ is given by
\cite{GROSS}
\beqa
g = S
\label{kgr}
\eeqa
Both $f$ and $g$ are universal, that is independent of the compactification
scheme used.  This model independence is due to the fact that the dilaton
supermultiplet arises in the spacetime sector of the world-sheet SCFT,
rather than in its internal sector.

The actual form of the holomorphic coupling functions (\ref{fab})
and (\ref{kgr}) is fixed by the result of particular on-shell
3- and 4-point string scattering amplitude calculations \cite{GROSS}.
The underlying idea of the S-matrix approach is to demand the tree level
Lagrangian (\ref{orblag}) to reproduce the on-shell tree string theory S-matrix
elements.  In such scattering amplitude calculations, contributions due to
the exchange of massless modes have to be extracted away.  The remaining
contributions due to the exchange of massive modes are organized in powers
of the string tension ${\alpha}'$.  In an on-shell scattering amplitude,
subtraction of the contribution from the massless exchanges doesn't
pose any problems, yielding unambiguous results such as (\ref{fab})
and (\ref{kgr}).  Going off-shell, however, introduces ambiguities in the
way of subtracting the massless contributions \cite{GROSS,TSEYT}, that
is, the subtraction scheme is not unique.  At the level of the tree level
Lagrangian (\ref{orblag}) these ambiguities translate into having undetermined
gravitational couplings $\Delta$ and $\Sigma$.  That is, in contrast to the
holomorphic functions $f_{ab}$ and $g$, $\Delta$ and $\Sigma$ cannot
be uniquely determined from the results of appropriate string scattering
amplitudes.  Now, the conventional choice \cite{GROSS,TSEYT} for the
gravitational couplings $\Delta$ and $\Sigma$ is
\beqa
\Delta = -  S \nonumber\\
\Sigma = 4  S
\label{choice}
\eeqa
Then, the terms in (\ref{orblag}) quadratic in the gravitational superfields
can be combined into the superfield expression (\ref{sgb}) containing the
Gauss-Bonnet combination.  The tree-level
Lagrangian (\ref{orblag}) is then rewritten
into
\beqa
\L &=& - \frac{3}{2} \kappa^{-2} \int d^4 \theta E \left[K\right]
+ \frac{1}{4} \int d^2 \Theta \;\; \epsilon
\; k_a S  \; W^{\alpha a} W_{\alpha a} \nonumber\\
&+& \frac{1}{4} \int d^2 \Theta \;\; \epsilon \;
S \;(8 W^{\alpha \beta \gamma}
W_{\alpha \beta \gamma} + ( {\cal D}^2 - 8 R)
(G^{\alpha \dot{\alpha}}
G_{\alpha \dot{\alpha}}
-4
{\bar R} R)) +  h.c.
\label{lgds}
\eeqa 
Thus, the conventional choice (\ref{choice}) for $\Delta$ and
$\Sigma$ assumes that there is a subtraction scheme which, in the tree
level Lagrangian (\ref{orblag}), translates into a coupling of the
dilaton supermultiplet $S$ to the super Gauss-Bonnet combination, only.

In fact, it has been argued that \cite{ZWIEB} the term quadratic in the
gravitational superfield $W^{\alpha \beta \gamma}
W_{\alpha \beta \gamma}$
may only appear in the super Gauss-Bonnet combination,
that is, the term in (\ref{orblag})
proportional to  $W^{\alpha \beta \gamma}
W_{\alpha \beta \gamma}$ always has to be completed to a super Gauss-Bonnet
combination.  Otherwise,
a naked $C^2$-term would appear which,
when added to the Einstein term $\cal{R}$, would lead to
modified equations of motion describing the propagation of
massive ghosts to gravity \cite{Stelle,FERN}.  Then, as
argued in \cite{ZWIEB},
the appearance of such ghost modes in string theory would violate unitarity
and, hence, naked $C^2$-terms shouldn't occur in effective string theory
Lagrangians.  It was pointed out in \cite{GROSS,TSEYT}, however, that since
these ghost modes are at the Planck mass, they
really are in a region of large momentum, for which the perturbative
$\alpha'$ expansion of the effective orbifold Lagrangian is unreliable.
Thus, there is a priori no imperative reason for the conventional
choice (\ref{choice}) of the gravitational couplings $\Delta$ and $\Sigma$.
On the other hand, one can generate
$G^{\alpha \dot{\alpha}}G_{\alpha \dot{\alpha}}$
and $\bar{R} R$ terms by field redefinition of the
dilaton multiplet.  This can be easily seen in the linear multiplet
representation of the dilaton multiplet.  In the linear
multiplet representation, the couplings of the dilaton multiplet $L$
are encoded in its modified Bianchi identity \cite{GRIMM1,GRIMM2}
\beqa
(\bar{\cal{D}}^2 - 8 \bar{R}) L = a W^{\alpha} W_{\alpha}
+ b W^{\alpha \beta \gamma} W_{\alpha \beta \gamma} +
(\bar{\cal{D}}^2 - 8 \bar{R}) (c G^{\alpha \dot{\alpha}}
G_{\alpha \dot{\alpha}} + d \bar{R} R)
\label{linear}
\eeqa
(\ref{linear}) clearly shows that the $G^{\alpha \dot{\alpha}}
G_{\alpha \dot{\alpha}}$ and the $\bar{R} R$ terms can be absorbed into
$L$ by a redefinition of the linear multiplet.
Thus, the conventional choice
(\ref{choice}) can be looked upon as a gauge choice \cite{TSEYT}, and we will
stick to it in the following.

Having determined the tree-level dependence of the gravitational
functions $g, \Delta$ and $\Sigma$, we now turn to the main issue of this
paper, namely to the moduli dependent
one-loop corrections to these gravitational coupling
functions.  These one-loop corrections arise in two categories.  There
are the finite moduli dependent threshold effects \cite{Dix,10}
associated with
integrating out all the massive
modes contained in the orbifold spectrum. But then, there are also quantum
effects due to the light modes in the theory, the massless particles,
giving rise to non-local terms in the effective orbifold Lagrangian.
When discussing the momentum dependence of physical
couplings all quantum corrections, massless and massive,
need to be taken into account.

We will, throughout this paper, assume that there is a regularisation
scheme which preserves local supersymmetry.  Then,
at low energies, $p^2 \ll  M^2_{String}$, the
one-loop threshold
contributions of the heavy modes
to the gravitational couplings $g, \Delta$ and $\Sigma$ are local
and parametrised as follows
\beqa
g(S,T) &=&  S + g^{1-loop}_H(T) \nonumber\\
\Delta(S,T,{\bar T}) &=& -  S + \Delta^{1-loop}_H(T,{\bar T}) \nonumber\\
\Sigma(S,T,{\bar T}) &=& 4 S + \Sigma^{1-loop}_H(T,{\bar T})
\label{loopcoup}
\eeqa
As the dilaton plays the role of a string-loop counting parameter,
it doesn't enter the 1-loop corrections to $g, \Delta$ and $\Sigma$
\cite{May2}.
It is important to note that
the requirement of $N=1$ spacetime supersymmetry restricts
the 1-loop correction to $g$ to be holomorphic, whereas the 1-loop
corrections to $\Delta$ and $\Sigma$ can, a priori, be non-holomorphic
functions of the moduli fields $T$.
Thus, it is important to emphasize that massive modes
can, in principle, contribute non-holomorphic terms at 1-loop to
some of the gravitational couplings, namely to $\Delta$ and to
$\Sigma$.  We will, in section 5, give a field theoretical explanation
for the possible appearance of such non-holomorphic terms.  Also note that
the 1-loop corrections to
$g, \Delta$ and $\Sigma$ vary from orbifold to orbifold, ruining
the tree-level universality of $g, \Delta$ and $\Sigma$.

Let us point out that, similarly to the case of the holomorphic
gauge coupling function $f_{ab}$, there is probably a non-renormalisation
theorem stating that the holomorphic gravitational coupling
function $g(S,T)$ doesn't receive corrections beyond one-loop.  Such a
non-renormalisation theorem probably doesn't apply to non-holomorphic
couplings such as $\Delta$ and $\Sigma$.

The moduli dependent
1-loop contributions to the gravitational couplings $g, \Delta$
and $\Sigma$
due to the massless modes can, on the other hand,
in principle be computed
by field theoretical means,
based on the knowledge of the
massless orbifold spectrum as well as on its
tree-level couplings,  as we will review in the next section.
These moduli dependent contributions to $g, \Delta$ and to $\Sigma$
are all non-local and, hence,  non-holomorphic.
Thus, it is important to notice
that the non-holomorphic 1-loop contributions to $g$ are due
only to the massless modes, whereas non-holomorphic contributions
to $\Delta$ and $\Sigma$ can arise both from massless and massive
contributions.

The effective gravitational couplings, derived from the associated
low energy effective Lagrangian of the orbifold theory, will contain
all of the above moduli dependent contributions due to both massless
and massive particles.
  Duality invariance of the effective gravitational
couplings will put constraints on the 1-loop moduli dependent
threshold corrections due to the massive modes.  This will be discussed
in section 5.

\section{Massless 1-loop contributions to gravitational couplings}

\setcounter{equation}{0}

\hspace*{.3in}  In this section, we will compute
the
moduli dependent 1-loop
contributions from the massless modes of the orbifold spectrum
to the gravitational coupling
functions $g, \Delta$ and $\Sigma$.
As pointed out in the previous section, these massless
contributions
can be computed in field theory
from the knowledge of the massless spectrum as well as of the
massless tree level orbifold Lagrangian.  A manifestly
supersymmetric procedure
for calculating these 1-loop contributions was presented in \cite{15}
and
consists in performing the calculations in the conventional
superspace formulation of supergravity-matter systems \cite{1,2}.  Let us
thus first review a few basic facts about conventional superspace
and then review the supersymmetric procedure introduced in \cite{15}.

The structure group of conventional superspace is taken
to be simply $SL(2,C)$ with the associated one-form gauge
connection $\phi_B\,^A = dz^M \phi_{MB}\,^A$.  In addition, one
introduces the supervielbein $E_M\,^A$ and the associated
one-forms $E^A = dz^M E_M\,^A$.  Solving the Bianchi identities
subject to a set of constraints \cite{16}, one finds that
all components of the torsion and curvature may be expressed in
terms of a set of superfields and their coordinate derivatives:
\beq {\rm superfield} \;\;\; R,\bar{R} \;\;\; G_{\a\dota} \;\;\;
W_{\a\b\gamma}, \;\; {\bar{W}}_{\dota \dot{\beta} \dot{\gamma}}
\eeq 
Since they are obtained by solving the Bianchi identities with
respect to a very different set of constraints, the $R$, $G_{\a
\dota}$ and $W_{\a \b \gamma}$ in this section are different
than, and not to be confused with, the field strength solutions
of the Bianchi identities in \Kahler\ superspace.  The relation
between them will be discussed later.

The tree-level kinetic
superfield \lag\ in this superspace
is given by
\beqa \L_0 = -3 \kappa^{-2} \int d^4 \theta E e^{-\frac{{\kappa}^2}{3}
K(\Phi, \bar{\Phi}) }
\label{kitl}
\eeqa 
Note that $E$ is the superdeterminant in conventional $SL(2,C)$
superspace and is not identical to the superdeterminant in
\Kahler\ superspace discussed earlier.  Now, the kinetic tree-level
\lag\ (\ref{kitl})
possesses, by construction, both gauge and Lorentz
invariance, as did the kinetic \Kahler\ superspace \lag
(\ref{lzero}).  However, it is
clearly not invariant under the \Kahler\ transformation
${\kappa}^2 K \ra {\kappa}^2 K + F +
\bar{F}$  (\ref{ktr}).  Instead of possessing pure K\"{a}hler symmetry,
it exhibits
a mixed super-Weyl-K\"{a}hler invariance,
as is well known
\cite{1,2,90}.  Super-Weyl transformations change the supervielbein
as follows \cite{90}
\beqa E_M \,^a &\ra& e^{\Sigma + \bar{\Sigma}} E_M\,^a \nonumber
\\ E_M\,^\a & \ra & e^{2 \bar{\Sigma} - \Sigma} \left( E_M\,^\a +
\frac{i}{2} E_M\,^b \left( \epsilon \sigma_b \right)^\a
\,_{\dota} \bar{\D}^{\dota} \bar{\Sigma} \right) \nonumber \\
E_{M \dota} &\ra& e^{2 \Sigma - \bar{\Sigma}} \left( E_{M \dota}
+ \frac{i}{2} E_M\,^b \left( \epsilon \bar{\sigma}_b
\right)_{\dota} \,^\a \;\D_\a \Sigma \right)
\label{superweyl}
\eeqa 
where $\Sigma$ and $\bar{\Sigma}$ are superfield parameters
subject to the chirality conditions
\beqa \bar{\D}^{\dota} \Sigma &=& 0 \nonumber \\
\D_\a \bar{\Sigma} &=& 0
\label{para}
\eeqa 
Under (\ref{superweyl})
\beq E \ra  e^{2 (\Sigma + \bar{\Sigma})} \; E
\label{sdet}
\eeq 
Chiral superfields and $K(\Phi, \bar{\Phi})$ are invariant
under super-Weyl transformations.  Clearly, the conventional kinetic
Lagrangian (\ref{kitl}) is left invariant under combined super-Weyl and
K\"{a}hler transformations, (\ref{ktr}) and
(\ref{superweyl}), provided
that
\beqa \Sigma &=& \frac{1}{6} F \nonumber \\ \bar{\Sigma} &=&
\frac{1}{6} \bar{F}
\label{swk}
\eeqa 
Inserting
K\"{a}hler potential (\ref{kpot})
for the orbifold theories under consideration
into the conventional tree-level
Lagrangian (\ref{kitl}) and expanding in powers of $\phi^i$ yields
\beqa \L_0 = - 3 \kappa^{-2} \int d^4 \theta
E e^{-\frac{\kappa^2}{3}  K_0}  + \sum_i
\int d^4 \theta E e^{-\frac{\kappa^2}{3} K_0}
{\cal Z}^i(T,\bar{T})
{\phi}^{\dagger i}  e^V \phi^i
+ \ldots \
\label{ctl}
\eeqa 
where
\beqa
{\cal Z}^i(T, {\bar T}) =
\prod_I
\left[ T^I + \bar{T}^I \right]^{q_I^i}
\label{zi}
\eeqa
We will, in the following, be interested in the moduli dependent
1-loop contributions of the fields $\phi^i$.  We will thus treat the
$\phi^i$ as quantum fields, whereas $E,V,T^I$ and $S$ will be treated
as classical background fields.  At the end of this section we will
generalize our results to also include 1-loop contributions from
the massless fields
$E,V,T^I$ and $S$.

The kinetic Lagrangian (\ref{ctl}) is invariant under mixed
super-Weyl-K\"{a}hler transformations of the background fields
$E$ and $K_0$.  Note that it is also invariant under conformal
transformations as follows.
These are defined by arbitrary super-Weyl transformations, as
given in (\ref{sdet}), accompanied by the superfield rescalings of the
quantum fields $\phi$ and $\bar{\phi}$, given by
\beqa \phi &\ra& e^{-2 \Sigma} \phi \nonumber \\
\bar{\phi} &\ra& e^{-2 \bar{\Sigma}} \bar{\phi}
\label{cfs}
\eeqa 
Before turning to the computation of
the 1-loop contributions, let us expand Lagrangian (\ref{ctl})
into component fields.
To lowest order in the background fields one obtains
\beqa \L_0/e &=&
\frac{1}{6}
\bar{A}^i A^i \left( \R - \frac{1}{2}
(4 ({\kappa}^2 K_0-3ln{\cal Z}^i) \left|_{\theta^2 \bar{\theta}^2}
+ \Box_0
({\kappa}^2 K_0-
3ln{\cal Z}^i) \left| ) \right) \right.  \right. \nonumber\\
&-& (1 - \frac{1}{3}(\kappa^2 K_0 - 3 ln {\cal Z}^i)|) \,\,
g^{mm} \tilde{\D}_m A^i \tilde{\D}_n \bar{A}^i \nonumber\\
&-&  (1-\frac{1}{3}(\kappa^2 K_0 - 3 ln {\cal Z}^i)|) \,\,
\frac{i}{2}  \left( \chi^i \sigma^m \tilde{\D}_m
\bar{\chi}^i - \left( \tilde{\D}_m \chi^i \right) \sigma^m \bar{\chi}^i
\right)
+ \ldots
\label{clag}
\eeqa 
where the covariant derivatives are
\beqa
\tilde{\D}_m A^i &=& ( \partial_m
-\frac{i}{3} \left( b_m - 2 i \kappa^2 a_m + 6 i {\cal Z}^i_m
\right) ) A^i
\nonumber \\
\tilde{\D}_m \chi^i &=&  ( \partial_m - \omega_m
 - \frac{2}{3}( {\kappa}^2 c_m - 3 {\cal Z}^i_m)) \chi^i
 \eeqa 
and where
\beqa c_m &=& a_m - {\kappa}^{-2} \frac{i}{4} b_m \nonumber\\
{\cal Z}^i_m &=& \frac{1}{4}(\partial_I ln{\cal Z}^i \partial_m T^I -
 \partial_{\bar I} ln{\cal Z}^i \partial_m {\bar T}^I)
\label{zmi}
\eeqa 
Note that $b_m, a_m, \omega_m$, $c_m$ and ${\cal Z}^i_m$ are all
evaluated at classical background field values.
The
combination $c_m$
appearing in the
covariant derivative of the Weyl fermion acts as a connection which insures
invariance under mixed super-Weyl-\Kahler\ transformations.  Under super
Weyl transformations it can be shown that
\beq  \delta b_m \left. = - 3 i \partial_m \left( \bar{\Sigma} -
\Sigma \right) \right| \eeq 
Then, under super Weyl-K\"{a}hler transformations (\ref{swk}), $c_m$
transforms as
\beq \left. \delta c_m = {\kappa}^{-2} \frac{3}{8}
\partial_m \left( F - \bar{F} \right) \right|
\label{dcm}
\eeq 
and the Weyl fermion $\chi^i$ as
\beq \delta \chi^i = - \left( 2 \bar{\Sigma} - \Sigma \right)
\left| \chi^i \right. \eeq 
with $\Sigma$ and $\bar{\Sigma}$ given as in (\ref{swk}).
Thus, it can be readily checked that the transformation (\ref{dcm})
of $c_m$ exactly cancels against the inhomogenous term occuring in
the transformation of the fermionic kinetic energy terms
under combined super-Weyl-K\"{a}hler
transformations.
Also
note that the combination $b_m - 2i \kappa^2 a_m$ appearing in
the covariant derivative $\tilde{\D}_m A$ is invariant
under mixed  super-Weyl-\Kahler\ transformations, as it must,
since the component scalar field $A$ does not transform under
mixed super-Weyl-\Kahler\ transformations.

We are now poised to compute the 1-loop moduli contributions due to
the massless fields $\phi^i$.  We will, for simplicity, in the following
only consider the contribution of one such field $\phi$.
As pointed out in \cite{15},
there are two
types of component graphs we will have to consider
in conventional superspace.  The first type of
component graphs will be related to the computation of the trace anomaly
in conventional superspace.  The second type of component graphs will be due
to mixed gravitational
anomalies involving the component connections $c_m$ and ${\cal Z}^i_m$.

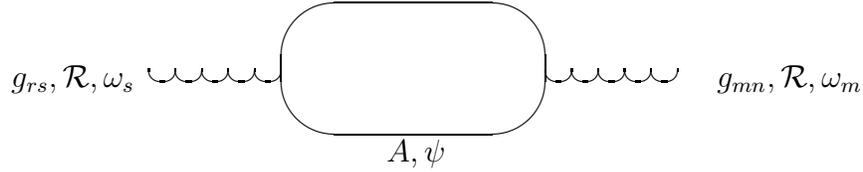
\begin{figure}[t]
\[
\begin{picture}(3,22)

\put(0,25){\oval(100,50)}
\multiput(55,25)(10,0){5}{\oval(10,10)[b]}
\put(115,18){$g_{mn},{\cal R},\omega_m$}
\multiput(-95,25)(10,0){5}{\oval(10,10)[b]}
\put(-152,18){$g_{rs},{\cal R},\omega_s$}
\put(-10,-10){$A, \psi$}
\end{picture}
\]
\caption{The gravitational two-point function}
\label{f1}
\end{figure}

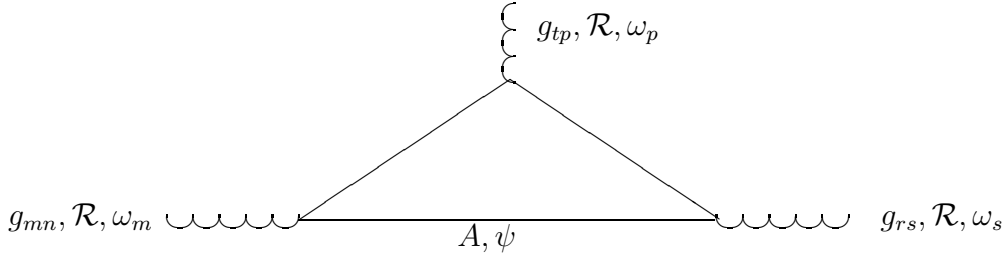
\begin{figure}[t]
\[
\begin{picture}(5,60)
 \put(-60,0){\line(10,0){157}}
 \multiput(-105,2)(10,0){5}{\oval(10,10)[b]}
\put(-60,0){\line(3,2){80}}
\put(20,53){\line(3,-2){79}}
\put(160,-2){$g_{rs},{\cal R},\omega_s$}
\put(-170,-2){$g_{mn},{\cal R},\omega_m$}
\multiput(103,2)(10,0){5}{\oval(10,10)[b]}
\multiput(22,57)(0,10){3}{\oval(10,10)[l]}
\put(30,70){$g_{tp},{\cal R},\omega_p$}
\put(0,-10){$A,\psi$}
\end{picture}
\]
\caption{The gravitational three-point function}
\label{f2}
\end{figure}

We begin with the component graphs relevant for the discussion of the
trace anomaly in conventional superspace.  These graphs are depicted
in Figure \ref{f1} and Figure \ref{f2}.
Prior to renormalisation these two graphs are, in
$4-2\varepsilon$ dimensions,
evaluated
to lowest order in the background fields to be \cite{30,29,Frad,33,19}
\beqa
\L_{unr} &=& \frac{1}{(4 \pi)^2} \; \frac{1}{48} \;
\frac{1}{\varepsilon}
\left(3 {\beta}_c
C_{mnpq} C^{mnpq} - {\beta}'_c GB \right) \nonumber\\
&-& \frac{1}{(4 \pi)^2} \; \frac{\beta_c}{144} \; {\cal R}^2 \; + \;
\frac{1}{(4 \pi)^2} \frac{1}{6 \cdot 24}
\left(3 {\beta}_c
C_{mnpq} C^{mnpq} - {\beta}'_c GB \right)
\frac{1}{\Box_0} \R \nonumber\\
&+& \cdots
\label{utm}
\eeqa 
where
\beqa {\beta}_c &=& \frac{1}{15} (N_S + 3 N_F)  \nonumber\\
{\beta}_c' &=& \frac{1}{15} (N_S + \frac{11}{2} N_F)
\label{betac}
\eeqa
$N_S(=2)$ is the number of real component scalars fields and $N_F(=1)$
is the number of component Weyl fermions.  Hence, $\beta_c=\frac{1}{3}$
and $\beta_c'=\frac{1}{2}$.  The
subscript $c$ indicates that these
are the
trace anomaly coefficients associated with a chiral supermultiplet.
The $\frac{1}{\varepsilon}$-contribution in (\ref{utm}) comes accompanied
by a $ln \Box_0$-piece, which we have dropped, since it
will be irrelevant in the following.
The dots in (\ref{utm}) stand for additional terms involving the
conformal scalar field $\Psi = 1-\frac{1}{6}(\Box +
\frac{1}{6} {\R})^{-1} \R$ of
Fradkin et Vilkovisky \cite{Frad}, whose role is to insure the vanishing
of the trace of the unrenormalised energy momentum tensor, as follows.
The energy momentum tensor is defined by
\beq T_{mn} = \frac{2}{\sqrt{-g}}\;\; \frac{\delta S}{\delta
g^{mn}}.\eeq 
Due to the conformal symmetry (\ref{cfs})
of the vertices in Lagrangian (\ref{clag}),
the trace of the
unrenormalised energy momentum tensor associated with (\ref{utm})
must vanish.  This is, in fact, the case, as can be checked by keeping
the following two things in mind.
First note that
the variation
of $\int \sqrt{-g} C^2_{mnpq}$ and of $\int \sqrt{-g} GB$
in $4-2\varepsilon$ dimensions are non-vanishing and
proportional to $\varepsilon$
\beqa
\frac{2 g^{mn}}{\sqrt{-g}} \frac{\delta}{\delta g^{mn}}
\int d^{4 - 2 \varepsilon}
x \sqrt{-g} GB &=& 2 \; \varepsilon \; GB \nonumber\\
\frac{2 g^{mn}}{\sqrt{-g}} \frac{\delta}{\delta g^{mn}}
\int d^{4 - 2\varepsilon}
 x \sqrt{-g} C^2 &=& 2 \; \varepsilon \;
(C^2  - \frac{2}{3} {\Box_0} {\cal R}) \nonumber\\
\frac{2 g^{mn}}{\sqrt{-g}} \frac{\delta}{\delta g^{mn}}
\int d^{4- 2 \varepsilon} x \sqrt{-g} {\cal R }^2 &=& - 12 \;
 \Box_0  {\cal R}
\label{var}
\eeqa
Secondly, also note that the
variation of the non-local term in (\ref{utm})
proportional to $GB$ yields not only a local term proportional to
$GB$, but also additional non-local terms.  These additional non-local
terms are precisely cancelled against the variation of the terms involving
the conformal scalar field $\Psi$ of Fradkin et al \cite{Frad}.

The renormalised trace, on the other hand, is
non-vanishing and evaluated to be
\beqa
T^{conf} \,^{m} \,_m = - \frac{1}{24} \; \frac{1}{(4\pi)^2} \left(
3{\beta}_c C_{mnpq} C^{mnpq} - {\beta}_c' GB  \right)
+ \frac{\beta_c}{12}\frac{1}{(4 \pi)^2} \; {\Box_0} {\cal R}
\label{renormtr}
\eeqa 
which is the well-known gravitational contribution to the
one-loop trace anomaly from conformal scalars and Weyl fermions \cite{29,33}.
Note that the role of the local ${\cal R}^2$-term in
(\ref{utm})
is solely to cancel against the ${\Box_0 {\cal R}}$-piece
in the variation (\ref{var}) of the $\frac{1}{\varepsilon} C^2$-term.
All local terms in (\ref{utm})
will be irrelevant in the subsequent discussion and so we
will drop them.

\begin{figure}[t]
\[
\begin{picture}(5,60)
 \put(-60,0){\line(10,0){157}}
 \multiput(-105,2)(10,0){5}{\oval(10,10)[b]}
\put(-60,0){\line(3,2){80}}
\put(20,53){\line(3,-2){79}}
\put(160,-2){$\omega_n$}
\put(-130,-2){$\omega_m$}
\multiput(103,2)(10,0){5}{\oval(10,10)[b]}
\multiput(20,52)(0,5){6}{\circle*{4}}
\put(30,60){$c_l$}
\put(0,-10){$\psi$}

\end{picture}
\]
\caption{Fermionic contribution to the mixed gravitational-K\"{a}hler
anomaly}
\label{f3}
\end{figure}
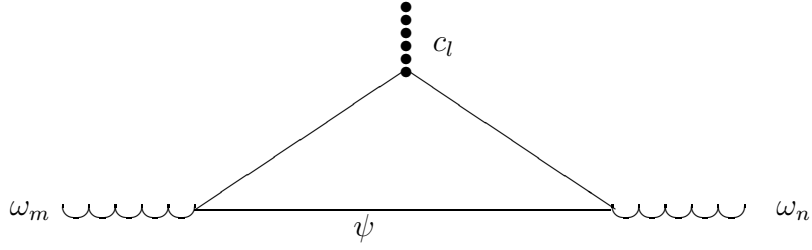

Next, let us look at the graph in Figure \ref{f3}.
It yields the fermionic contribution to the mixed chiral
super-Weyl-K\"{a}hler-Lorentz anomaly.  This graph can be evaluated \cite{17}
and the associated effective Lagrangian is found to be
\beqa \L'_\chi = - \frac{2i{\kappa}^2}{3} \;\; \frac{1}{24} \;\;
\frac{1}{(4\pi)^2} \R_{mn a}\,^b \tilde{\R}^{mn}\,_b\,^a
\frac{1}{\Box_0} \partial^p c_p
\label{cmanom}
\eeqa 
where, again, $c_m= a_m - {\kappa}^{-2} \frac{i}{4} b_m$.
It is readily seen that (\ref{cmanom}) is anomalous under
super-Weyl-K\"{a}hler transformations (\ref{dcm}).  Adding the CP-odd
$b_m$-
term in (\ref{cmanom}) to the CP-even trace anomaly
contribution (\ref{utm}) yields
\beqa
\L^{conf}
&=& - \frac{2}{12 \cdot 24 \cdot (4 \pi)^2}
\R_{mna}\,^b \; \tilde{\R}^{mn\;\;a} \!\!\!_b \frac{1}{\Box_0}
\partial^p b_p \nonumber \\
&& + \frac{1}{12 \cdot 24 \cdot (4\pi)^2} C_{mnpq} C^{mnpq}
\frac{1}{\Box_0} 6{\beta}_c \R          \nonumber \\
&&- \frac{1}{12 \cdot 24 \cdot (4\pi)^2}
GB \frac{1}{\Box_0}
 2{\beta}_c' \R + \cdots
\label{lconf}
\eeqa 
where the dots stand again for the additional terms
involving Fradkin's et al conformal scalar field $\Psi$.
Using the identities \cite{25}
\beqa \frac{1}{16} (C_{mnpq} C^{mnpq} - i
\R_{mna}\,^b \; \tilde{\R}^{mn\;\;a} \!\!\!_b ) =
W^2_{\alpha \beta \gamma} |_{{\theta}^2} \nonumber \\
GB = \left(
8 W^2_{\a\b\gamma} + \left( \bar{\D}^2 - 8 R\right) \left(
G^2_{\a\dota} - 4 \bar{R} R \right) \right)|_{{\theta}^2} + h.c.
\eeqa 
it follows that (\ref{lconf}) is the component field expansion of the
following superfield Lagrangian
\beqa \L^{conf} &=& \frac{2}{3 \cdot (4\pi)^2} \int d^4 \theta
W^2_{\a\b\gamma}
\frac{1}{\Box_0}  (3{\beta}_c-{\beta}'_c) \bar{R}
\nonumber \\
&-& \frac{1}{12 \cdot (4\pi)^2} \int d^4 \theta
\left( \bar{\D}^2 - 8 R\right) \left(
G^2_{\a\dota} - 4 \bar{R} R \right)
\frac{1}{\Box_0} {\beta}'_c \bar{R} + \cdots
+ h.c.
\label{supertrace}
\eeqa 
The dots stand now for the terms involving the supersymmetric analogue
$\Upsilon = 1 + \Box_{-}^{-1} ({\bar{\cal D}}^2 - 8 R) \bar{R}$
\cite{19,Buch} of the conformal scalar field $\Psi$.  These terms are
given in \cite{Buch}.

\begin{figure}[t]
\[
\begin{picture}(5,60)
 \put(-60,0){\line(10,0){157}}
 \multiput(-105,2)(10,0){5}{\oval(10,10)[b]}
\put(-60,0){\line(3,2){80}}
\put(20,53){\line(3,-2){79}}
\put(160,-2){$E$}
\put(-130,-2){$E$}
\multiput(103,2)(10,0){5}{\oval(10,10)[b]}
\multiput(20,53)(0,5){6}{\circle{4}}
\put(30,60){$K_0$}
\put(10,-10){$\phi$}

\end{picture}
\]
\caption{The anomalous mixed supergravity-K\"{a}hler-supergraph}
\label{f4}
\end{figure}
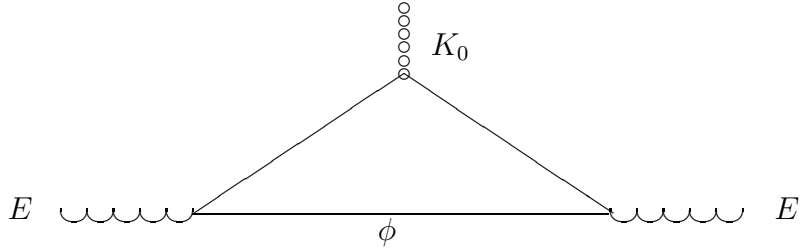

Now consider the term in (\ref{cmanom}) involving the
K\"{a}hler connection $a_m$.  Such a
component
term is uniquely contained in the
following superfield expression
\beqa \tilde{{\L}}_{K} &=& - \frac{2}{3 \cdot 24 \cdot
(4\pi)^2} \int d^4 \theta
W^2_{\a\b\gamma}
\frac{1}{\Box_0} \left( \kappa^2
D^2 K_0
\right) + h.c.
\label{wabc}
\eeqa
This superfield expression, on the other hand, arises when evaluating
the supergraph depicted in Figure \ref{f4}.  The result of such a supergraph
calculation might, however, also yield additional terms proportional
to
$G^{\alpha \dot{\alpha}} G_{\alpha \dot{\alpha}}$ and $\bar{R} R$.
$G^{\alpha \dot{\alpha}} G_{\alpha \dot{\alpha}}$ and $\bar{R} R$
contain only CP-even terms in their highest components, that is, they
do not contain
$\tilde{\cal R} {\cal R}$ in their component expansion and, thus,
they do not spoil the component result (\ref{cmanom}).
Then, the actual result of the supergraph calculation should read
\beqa \L_K &=& - \frac{2}{3 \cdot 24 \cdot(4\pi)^2} \int d^4 \theta
W^2_{\a\b\gamma}
\frac{1}{\Box_0} \left( \kappa^2
D^2 K_0
\right) \nonumber \\
&+& \frac{\delta_c}{12 \cdot 24 \cdot (4\pi)^2} \int d^4 \theta
\left( \bar{\D}^2 - 8 R\right) \left(
G^2_{\a\dota} - 4 \bar{R} R \right)
\frac{1}{\Box_0} \left(  \kappa^2 D^2 K_0 \right) \nonumber \\
&+& \frac{\zeta_c}{12 \cdot 24 \cdot (4\pi)^2} \int d^4 \theta
\left( \bar{\D}^2 - 8 R\right) \bar{R} R
\frac{1}{\Box_0} \left(  \kappa^2 D^2 K_0 \right) \nonumber \\
&+& h.c.
\label{wgr}
\eeqa 
where $\delta_c$ and $\zeta_c$ denote the two coefficients which we
haven't computed.  One can, however, argue that the two unknown coefficients
$\delta_c$ and $\zeta_c$ have to be zero, as follows.
Since the supergraph in Figure \ref{f4} is constructed out of conformal
vertices only, one expects that the resulting expression should respect
the conformal symmetry (\ref{cfs}).  Since
the only superfield expression allowed by the conformal
symmetry (\ref{cfs}) is $W_{\alpha
\beta \gamma}^2$, one expects the result of this supergraph
calculation to be given by (\ref{wabc}), only.  Hence, we will
in the following set $\delta_c = \zeta_c = 0$.  It should also be pointed
out that the conclusions contained in this paper remain valid
even for
non-vanishing
$\delta_c$ and $\zeta_c$.

Adding up (\ref{supertrace}) and (\ref{wabc}) yields
the supersymmetric mixed gravitational
super-Weyl-K\"{a}hler anomaly as
\beqa \L &=& \frac{2}{3 \cdot (4\pi)^2} \int d^4 \theta
W^2_{\a\b\gamma}
\frac{1}{\Box_0} \left( (3{\beta}_c-{\beta}'_c)
\bar{R} - \frac{\kappa^2}{24}
D^2 K_0
\right) \nonumber \\
&-& \frac{1}{12 \cdot (4\pi)^2} \int d^4 \theta
\left( \bar{\D}^2 - 8 R\right) \left(
G^2_{\a\dota} - 4 \bar{R} R \right)
\frac{1}{\Box_0}
{\beta}'_c \bar{R}
\nonumber \\
&+& \cdots + h.c.
\label{swka}
\eeqa 
Its variation under combined super-Weyl-K\"{a}hler transformations
\beqa
&&\kappa^2 \delta K = F + \bar{F} \nonumber\\
&&\delta \bar{R} = - \frac{1}{24} D^2 F
\eeqa
can be computed using the techniques in \cite{Buch}, and it is
found to be
\beqa
\delta {\cal L} &=& \frac{1}{72 \cdot (4 \pi)^2}
\int d^2 \Theta \left[ 8 ( 1 + 3 \beta_c - \beta_c') W_{\alpha
\beta \gamma}^2 -
\beta_c'
 (\bar{\cal D}^2 - 8 R) ( G_{\alpha
\dot{\alpha}} - 4 \bar{R} R )
\right] \; F \nonumber\\
&+&  h.c.
\label{susytr}
\eeqa

We now turn to the mixed gravitational-$\sigma$-model anomaly
which
arises through the coupling of the fermionic current to the
connection ${\cal Z}_m$ given in (\ref{zmi}).
The associated anomalous component graph is the same as in Figure \ref{f3}
with $c_m$ replaced by $-3 {\cal Z}_m$.  The associated anomalous supergraph
is the same as in Figure 4 with $K_0$ replaced by $-3ln{\cal Z}$, where
${\cal Z}$ is given by (\ref{zi}).  Thus, the above result (\ref{swka})
can be readily extended to also include the contribution from the
mixed gravitational $\sigma$-model anomaly.
Collecting all
the terms of interest and generalising the above results to also
include any
number of fields $\phi^i$ yields
\beqa \L &=& \sum_i \left[
\frac{2}{3 \cdot (4\pi)^2} \int d^4 \theta
W^2_{\a\b\gamma}
\frac{1}{\Box_0} \left( (3{\beta}_c-{\beta}'_c)
\bar{R} - \frac{1}{24}
D^2 (\kappa^2 K_0 -3ln{\cal Z}^i)
\right) \right. \nonumber \\
&-& \frac{1}{12 \cdot (4\pi)^2} \int d^4 \theta
\left( \bar{\D}^2 - 8 R\right) \left(
G^2_{\a\dota} - 4 \bar{R} R \right)
\frac{1}{\Box_0}
{\beta}'_c \bar{R}
\nonumber \\
&+& \left. \cdots + h.c. \right]
\label{phiconv}
\eeqa 

Lagrangian (\ref{phiconv})
contains the non-local 1-loop contributions to the gravitational
couplings due to massless $\phi^i$ fields in conventional superspace.
We would now like to transform it over into K\"{a}hler superspace using
the manifestly supersymmetric procedure introduced in  \cite{6}.
Both formulations of
supergravity-matter are related by particular
superfield rescalings of the
underlying torsion constraints \cite{6}.  This implies,
among other things, that the superfields $R$ and $G_{\alpha \dot{\alpha}}$
of the conventional superspace formulation are related to the
superfields $R$ and $G_{\alpha \dot{\alpha}}$
of the \kahler\ superspace
formulation by \cite{6}
\beqa R &\ra R & -  \frac{\kappa^2}{24} {\bar{D}}^2 K \nonumber\\
G_{\alpha \dot{\alpha}} &\ra & G_{\alpha \dot{\alpha}}  -
\frac{\kappa^2}{12} [ D_{\alpha}, \bar{D}_{\dot{\alpha}}] K
\label{rtrans}
\eeqa
where we work to linearised level in $K$ only.
Applying (\ref{rtrans}) to Lagrangian (\ref{phiconv}) yields
the supersymmetric mixed gravitational-K\"{a}hler and mixed
gravitational-$\sigma$-model anomalies in K\"{a}hler superspace as
\beqa
\L_c^{anom} &=& \sum_i \left[
\frac{1}{24 \cdot(4\pi)^2} \int d^4\theta \left[
\frac{2}{3}(1+3{\beta}_c-{\beta}'_c)W^2_{\alpha\beta\gamma}
\right. \right. \nonumber \\
&-& \left. \left.
\frac{\beta'_c}{12}\,
(\bar \D^2-8R)
(G^2_{\alpha \dot\alpha} - 4\bar{R} R)\right] \frac{1}{\Box_0}D^2
\left( - \kappa^2 K_0 \right) \right. \nonumber \\
&+& \frac{1}{24 \cdot(4\pi)^2} \int d^4\theta
W^2_{\alpha\beta\gamma}
\frac{1}{\Box_0}D^2
\left( 2 ln{\cal Z}^i  \right)
+ h.c.  \left. \right]
\label{phika}
\eeqa 
where we have added subscript $c$ to ${\cal L}^{anom}$ to
indicate that this
is the result associated with chiral supermultiplets $\phi$.
Using that $1+ 3\beta_c - \beta'_c = \frac{3}{2}$ and that
$\beta'_c=\frac{1}{2}$,
it follows that Lagrangian (\ref{phika}) can be rewritten as
\beqa
\L_c^{anom} &=& \sum_i \left[
\frac{1}{24 \cdot(4\pi)^2} \int d^4\theta
W^2_{\alpha\beta\gamma}
\frac{1}{\Box_0}D^2
\left( - \kappa^2 K_0 + 2 ln{\cal Z}^i  \right) \right.  \nonumber \\
&-& \frac{1}{24 \cdot 24 \cdot(4\pi)^2} \int d^4\theta
(\bar \D^2-8R)
(G^2_{\alpha \dot\alpha} - 4 \bar{R} R) \frac{1}{\Box_0}D^2
\left( - \kappa^2 K_0 \right) \nonumber\\
 &+& h.c.  \left. \right]
\label{lagcanom}
\eeqa
Note that the coefficient of $W^2_{\alpha \beta \gamma}$ is given
by $\kappa^2 K_0-2 ln{\cal Z}^i$ and, thus, is the correct coefficient.
Varying (\ref{lagcanom}) with respect to K\"{a}hler transformations
(\ref{ktr}) yields
\beqa
\delta_c^{anom} {\cal L} &=&
\sum_i \frac{1}{6 \cdot (4 \pi)^2} \int d^2 \Theta \left[
W_{\alpha \beta \gamma}^2 - \frac{1
}{24} (\bar{{\cal D}^2} - 8 R)
(G_{\alpha \dot{\alpha}}^2 - 4 {\bar R}R)
 \right] F + h.c.
\label{ksusytr}
\eeqa
which reproduces (\ref{susytr}).

It
is now not
difficult to extend the above results to include gauge vector multiplets
running around the
internal loop \cite{15}.  The only graphs occuring now are the trace anomaly
graphs (Figures 1 and 2) with a vector multiplet running in the loop.
All other graphs are absent.  Thus, the only
changes are that the $\beta$ and $\beta'$ coefficients
are now to be
evaluated for gauge vector multiplets.
Also note that,
since the
sigma-model Christoffel connection does not couple to gauginos, the $ln
{\cal Z}^i$
terms do not appear.  The anomalous Lagrangian associated with vector
supermultiplets
running around the loop is then given by
\beqa {\cal L}^{anom}_v &=&
\frac{dim \; G}{24
\cdot(4\pi)^2} \int d^4 \theta \left[
\frac{2}{3}(3 \beta_v- \beta'_v)W^2_{\alpha\beta\gamma} -
\frac{\beta'_v}{12} \right. \nonumber \\ &&\left. \times (\bar {\cal D}^2 -
8R)(G^2_{\alpha\dot\alpha} - 4\bar{R} R)\right]  \frac{1}{\Box_0}\;
D^2\left(-\kappa^2 \; K_0 \right) +
h.c.
\label{vecka}
\eeqa 
where $\beta_v$ and $\beta'_v$ are the trace anomaly coefficients of
$(C_{mnpq})^2$ and
$GB$ respectively due to a single vector supermultiplet given by
 \beqa \beta_v & = &
\frac{1}{15}(3N_F + 12 N_V) \nonumber \\ \beta'_v &=& \frac{1}{15} \left(
\frac{11}{2}
N_F + 62 N_V \right)
\eeqa 
 where $N_F(=1)$ is the number of component Weyl fermions and
$N_V(=1)$ is the number of component vector fields.
${\rm dim}G$ denotes the dimension of the gauge group.
Inserting $\beta_v=1$ and $\beta'_v=\frac{9}{2}$ into (\ref{vecka})
yields
\beqa {\cal L}^{anom}_v &=&
\frac{dim \; G}{24
\cdot(4\pi)^2} \int d^4 \theta \left[
 - W^2_{\alpha\beta\gamma} -
\frac{3}{8} \right. \nonumber \\ &&\left. \times (\bar {\cal D}^2 -
8R)(G^2_{\alpha\dot\alpha} - 4 \bar{R} R)\right]  \frac{1}{\Box_0}\;
D^2\left(-\kappa^2  \; K_0 \right)   +
h.c.
\label{vectorka}
\eeqa
Note that Lagrangian (\ref{vectorka}) reproduces the correct
anomaly coeficient of $W^2_{\alpha \beta \gamma}$.

We now turn to the contributions steming from moduli fields running
in the loop.
In order to calculate these contributions, we first need to expand the
K\"{a}hler potential $K_0$ (\ref{kts})
in powers of moduli quantum fluctuations
$\delta T^I$ as
\beqa
K_0 &\ra& K_0 + \sum_I \left[ - \frac{\delta T^I}{T^I + \bar{T}^I}
- \frac{\delta \bar{T}^I}{T^I + \bar{T}^I}  \right.  \nonumber\\
&+& \left. \frac{\delta T^I \delta \bar{T}^I}{(T^I + \bar{T}^I)^2}
+ \frac{1}{2} \frac{(\delta T^I)^2}{(T^I + \bar{T}^I)^2}
+ \frac{1}{2} \frac{(\delta \bar{T}^I)^2}{(T^I + \bar{T}^I)^2}
+ \dots
\right]
\label{qfdt}
\eeqa
where $K_0$ is again evaluated at the classical background.
Inserting (\ref{qfdt}) into the tree level Lagrangian (\ref{ctl}) and
keeping all terms up to quadratic order in $\delta T^I$ yields
\beqa \L_0 &=& - 3 \kappa^{-2} \int d^4 \theta
E e^{-\frac{\kappa^2}{3}  K_0}  + \frac{2}{3}
\int d^4 \theta E e^{-\frac{\kappa^2}{3} K_0} \left[
{\cal Z}^I \delta T^I \delta \bar{T}^I  \right. \nonumber\\
&+& \left.  \frac{1}{2} {\cal Z}^I (\delta T^I)^2
+ \frac{1}{2} {\cal Z}^I (\delta \bar{T}^I)^2 + \dots \right]
\label{lagdt}
\eeqa
where
\beqa
{\cal Z}^I(T, {\bar T}) =
\left[ T^I + \bar{T}^I \right]^{ q^I}  \;\;\; q^I= - 2
\label{zt}
\eeqa
Note that the linear terms in $\delta T^I$ and $\delta \bar{T}^I$
do not occur in (\ref{lagdt}) as they vanish due to the equation of
motion of the moduli fields.  Also note that the vertices in (\ref{lagdt})
occur in two types.  The $\delta T \delta \bar{T}$-vertices are of the
conformal type (\ref{cfs}), whereas the $(\delta T)^2$-vertices are not.
The vertices of the conformal type will yield 1-loop contributions similar
to the ones in (\ref{lagcanom}).  The conformal breaking vertices will
contribute additional non-conformally invariant terms proportional
to $\bar{R} R$.  This can be seen
as follows.  The component field expansion of the $(\delta T)^2$-term
in (\ref{lagdt}) yields, among other things, a term proportional to
${\cal R} A^2$.  Such an additional non-conformally invariant coupling
leads to an additional ${\cal R}^2$-term in the
trace of the renormalised stress energy momentum tensor (\ref{renormtr}).
Consequently, the super-Weyl-K\"{a}hler anomaly (\ref{susytr}) will
contain an additional $\bar{R} R$-contribution.  The corresponding K\"{a}hler
anomaly in K\"{a}hler superspace (\ref{ksusytr})
will also contain an additional
term proportional to $\bar{R} R$.  We will denote this contribution
by $\gamma_T$.  There is also an additional $\bar{R} R$-contribution
to the mixed gravitational-$\sigma$-model anomaly, which we will denote by
$\upsilon_T$.
Then, the moduli field contributions read
\beqa
\L_c^{anom} &=& \sum_I \left[
\frac{1}{24 \cdot(4\pi)^2} \int d^4\theta
W^2_{\alpha\beta\gamma}
\frac{1}{\Box_0}D^2
\left( - \kappa^2 K_0 + 2 ln{\cal Z}^I  \right) \right.  \nonumber \\
&-& \frac{1}
{24 \cdot 24 \cdot(4\pi)^2} \int d^4\theta
(\bar \D^2-8R)
(G^2_{\alpha \dot\alpha} - 4\bar{R} R) \frac{1}{\Box_0}D^2
\left( - \kappa^2 K_0 \right) \nonumber \\
&+& \left.  \frac{1}{12 \cdot 24 \cdot(4\pi)^2} \int d^4\theta
 (\bar \D^2-8R)
(\bar{R} R) \; \frac{1}{\Box_0}D^2
\left(
- \gamma_T \, \kappa^2 K_0 + 3
\upsilon_T
\,  ln{\cal Z}^I \right) \right. \nonumber\\
&+& \left.  h.c.   \right]
\label{modka}
\eeqa

Finally, let us discuss the 1-loop moduli dependent contributions
due to the supergravity multiplet and the dilaton multiplet.  First note
that both of them are present in any orbifold theory, and that their
contributions are the same from orbifold to orbifold.  Thus, their
contributions need not be treated separately, but rather can be combined
together.  Again, the associated supergraphs will not
be of the conformal type.
Consequently, their moduli dependent contributions, which we haven't
computed, can be parametrised
as follows
\beqa \L^{E,S} &=& \frac{21 + 1}{24 \cdot (4\pi)^2} \int d^4 \theta
W^2_{\a\b\gamma}
\frac{1}{\Box_0} \left( - \kappa^2
D^2 K_0
\right) \nonumber \\
&+& \frac{\xi}{24 \cdot (4\pi)^2} \int d^4 \theta
\left( \bar{\D}^2 - 8 R\right) \left(
G^2_{\a\dota} - 4 {\bar R} R\right)
\frac{1}{\Box_0} \left( - \kappa^2 D^2 K_0 \right) \nonumber\\
&+& \frac{\varrho}{12 \cdot 24 \cdot (4\pi)^2} \int d^4 \theta
\left( \bar{\D}^2 - 8 R\right) \left(
\bar{R} R \right)
\frac{1}{\Box_0} \left( - \kappa^2 D^2 K_0 \right) + h.c. \eeqa 
Note that the coefficient of $W^2_{\alpha \beta \gamma}$ is
known \cite{19} and
given
by the contributions of the gravitino (21) and the dilatino (1) to the
mixed gravitational-K\"{a}hler anomaly in K\"{a}hler superspace.
The unknown coeficients $\xi$ and $\varrho$ stand for additional
CP even terms proportional to $(G^2_{\alpha \dot{\alpha}} - 4{\bar R}R)$
and to $\bar{R} R$,
respectively.

Combining all the contributions from the massless orbifold fields,
from $\phi^i, T^I, V, E$ and the dilaton multiplet, and inserting
the $\sigma$-model metrics ${\cal Z}^i$ and ${\cal Z}^I$ into
the mixed anomalies yields the following total massless result
\beqa \L^{massless} &=& \sum_I \left[
\frac{b^I}{24 \cdot (4\pi)^2} \int d^4 \theta
W^2_{\a\b\gamma}
\frac{1}{\Box_0}
D^2 ln(T^I + {\bar T}^I)
  \right.  \nonumber \\
&+& \frac{p^I}{24 \cdot (4\pi)^2} \int d^4 \theta
\left( \bar{\D}^2 - 8 R\right) \left(
G^2_{\a\dota} - 4 {\bar R} R \right)
\frac{1}{\Box_0}  D^2 ln(T^I + {\bar T}^I)  \nonumber\\
&+& \left. \frac{h^I}{24 \cdot (4\pi)^2} \int d^4 \theta
\left( \bar{\D}^2 - 8 R\right) \left(
\bar{R} R \right)
\frac{1}{\Box_0} D^2 ln(T^I + {\bar T}^I)
+ h.c. \right]
\label{lagmless}
\eeqa 
The coefficient $b^I$ is completely known and given by \cite{40}
\beqa
b^I = 21 + 1 + n_M^I - {\rm dim}G + \sum_i (1+2q^i_I)
\label{ocoef}
\eeqa
where $n_M^I$ denotes the contribution from the moduli $T^I$
given by $n_M^I = 3 + 2 q^I = -1$.
The coefficients $p^I$ and $h^I$ are given by
\beqa
p^I &=& -\frac{3}{8} dim \; G - \frac{1}{8}
  - \frac{1}{24} (\sum_i 1)
    + \xi \nonumber\\
h^I &=& \frac{1}{12} [ 3 \gamma_T
+
3 \upsilon_T
q^I
 + \varrho ]
\label{hcoeff}
\eeqa
and, thus, determined by the unknown coeficients $\gamma_T$,
$\upsilon_T$, $\xi$
and $\varrho$.  Lagrangian (\ref{lagmless}) can be rewritten into
\beqa \L^{massless} &=& \sum_I \left[
\frac{b^I - 8 p^I}{24 \cdot (4\pi)^2} \int d^4 \theta
W^2_{\a\b\gamma}
\frac{1}{\Box_0}
D^2 ln(T^I + {\bar T}^I) \right.  \nonumber \\
&+& \frac{p^I}{24 \cdot (4\pi)^2} \int d^4 \theta
\left( 8 W^2_{\a\b\gamma} + (\bar{\D}^2 - 8 R) \left(
G^2_{\a\dota} - 4 {\bar R} R \right)\right)
\frac{1}{\Box_0}  D^2 ln(T^I + {\bar T}^I)  \nonumber\\
&+& \left. \frac{h^I}{24 \cdot (4\pi)^2} \int d^4 \theta
\left( \bar{\D}^2 - 8 R\right) \left(
\bar{R} R \right)
\frac{1}{\Box_0}  D^2 ln(T^I + {\bar T}^I)
+ h.c. \right]
\label{lagmlessgb}
\eeqa 
This shows that
the non-vanishing of $p^I$ results into a non-vanishing
Gauss-Bonnet contribution of amount $p^I$, yielding a leftover naked
$C^2$-term of amount $b^I-8p^I$.
The coefficient $h^I$, on the other hand,
determines the amount of an naked ${\cal R}^2$-term due to the massless
modes.  Note that $b^I$, $p^I$ and $h^I$ vary from orbifold to orbifold,
as they depend on the number of moduli and matter fields involved.

Let us now also take into account the Green-Schwarz mechanism which removes
some of the above contributions \cite{8,DFKZ}.  The Green-Schwarz mechanism
makes use of two ingredients, namely the Green-Schwarz term and the tree
level couplings of the dilaton multiplet to quadratic gravitational curvature
terms as given in (\ref{lgds}).  As stated in section 3, (\ref{lgds})
reflects the conventional choice of the tree level couplings of the
dilaton and pressuposes that, in the S-matrix approach, there is a
subtraction scheme which results in (\ref{kgr}) and (\ref{choice}).  As
mentioned in section 3, we will throughout this paper stick to that choice.
Then, the Green-Schwarz mechanism only affects the Gauss-Bonnet contribution
and it removes an amount $\delta^I_{GS}$ from the Gauss-Bonnet contribution
in (\ref{lagmlessgb}).  The new coefficients
$b^I$ and $p^I$ in
(\ref{lagmless}) then read \cite{40}
\beqa
b^I &=& 21 + 1 + n_M^I - {\rm dim}G + \sum_i (1+2q^i_I)
- 24\; \delta^I_{GS}  \nonumber\\
p^I &=& -\frac{3}{8} dim \; G - \frac{1}{8}
 - \frac{1}{24} ( \sum_i 1)
  + \xi
 - 3 \; \delta^I_{GS}
\label{ncoef}
\eeqa
The coefficients $\delta^I_{GS}$ determine the Green-Schwarz term
\cite{8,DFKZ}
\beqa
\L_{GS} = \int d^4 \theta E L\; \delta^I_{GS} \; ln(T^I + \bar{T}^I)
\eeqa
For orbifolds with $N=2$ sectors, some of the coefficients
$b^I$ (\ref{ncoef}) are non-vanishing \cite{13}.  We will, in Appendix A,
consider an example of such an orbifold with $N=2$ sectors, namely
$Z_4$.  There, we will show
that the field theoretical
calculation of the
coefficient $b^3$, associated with a complex plane for which there
is an $N=2$ sector,
agrees with the string scattering
amplitude calculation of \cite{13}.  Finally, we show in Appendix B
that the coefficients $b^I-8p^I$ of the naked $C^2$-term cannot be
all set
to zero
in the $Z_4$ orbifold by an appropriate
choice of the unknown coefficient
 $\xi$.
  Thus, assuming that
there is no Green-Schwarz removal other than of a Gauss-Bonnet combination,
it appears that the massless sector of an orbifold theory will in general
contribute naked $C^2$-terms to gravitational couplings.

\section{Massive 1-loop contributions to gravitational couplings}

\setcounter{equation}{0}

\hspace*{.3in}  In this section,
we will consider a field theoretical
toy model consisting of a massive chiral superfield
with a moduli dependent mass, and parametrize its 1-loop moduli
contributions to the gravitational couplings in a suitable way.  We then
turn to string theory and, much in the same spirit, introduce a suitable
parametrisation of the 1-loop contributions of the infinite tower
of massive string states.

Let us consider the following toy model in
conventional superspace
\beqa \L_0 &=& - 3 \kappa^{-2} \int d^4 \theta
E e^{-\frac{\kappa^2}{3}  K_0({T,\bar{T}})}  +
\int d^4 \theta E e^{-\frac{\kappa^2}{3} K_0({T,\bar{T}})}
{\cal Z}(T,\bar{T})
\bar{\phi}  \phi \nonumber\\
&+&  \frac{1}{2} \,\int d^2 \Theta \epsilon {\cal M}(T) \phi^2
+ \frac{1}{2} \,  \int d^2 \bar{\Theta} \bar{\epsilon}
\bar{{\cal M}}(\bar{T}) \bar{\phi}^2
+ \ldots
\label{toy}
\eeqa
where
\beqa
K_0 &=& -\sum_I ln(T^I + \bar{T^I}) \nonumber\\
{\cal Z} &=& \sum_I (T^I + \bar{T^I})^{q^I}
\eeqa
Lagrangian (\ref{toy}) describes the
coupling of a massive chiral superfield $\phi$ to supergravity in
conventional superspace.  Let us display some of the relevant bosonic
component terms in (\ref{toy}).  To lowest order in the background fields
one obtains
\beqa \L_0/e &=& - (1-\frac{1}{3}(\kappa^2 K_0 - 3 ln {\cal Z})|) \;
g^{mm} {\partial}_m A {\partial}_n \bar{A}
\nonumber\\
&+&
\frac{1}{6}
\bar{A} A \left( \R - \frac{1}{2}
(4 (\kappa^2 K_0-3ln{\cal Z}) \left|_{\theta^2 \bar{\theta}^2}
+ \Box_0
(\kappa^2 K_0-3ln{\cal Z}) \left| ) \right)  \right. \right. \nonumber\\
&+& (1- \frac{1}{3} ( \kappa^2 K_0 - 3 ln {\cal Z} )|)
\, {\cal F} {\bar{\cal F}} +
m(T)  A {\cal F} + \bar{m}(\bar{T})  \bar{A} \bar{\cal F} +
\ldots
\label{massl}
\eeqa 
where $m(T)= {\cal M}(T)|$ has mass dimension 1.  Similarly to
the massless case, one again has to
consider
two types of graphs when calculating the 1-loop moduli dependent contributions
from the massive superfield $\phi$
to quadratic gravitational couplings.  The first type of graphs consists of
the ones shown in Figures \ref{f1} and \ref{f2}, but this time with a
massive field running in the loop.  The second type consists of the graph
depicted in Figure \ref{f4} with $\kappa^2 K_0$ replaced by $\kappa^2
K_0 - 3 ln {\cal Z}$, again with a massive chiral superfield $\phi$
running in the loop.
We will, in the following, work
in a low-energy regime, i.e. at
$p^2 \ll |m(\langle T \rangle)|^2$.
Here, we have expanded $T$ around a vev $\langle T \rangle$.
In order to get a flavour for the terms appearing
in the calculation of these
graphs, we will in the following
compute some of them.

First
consider
the two-point function shown in Figure \ref{f1}
with two ${\cal R}$-legs sticking out
and scalar field $A$ running in the loop.
The relevant vertices are given by the term $\frac{1}{6} \bar{A} A {\cal R}$
in (\ref{massl}).  This particular graph can be readily evaluated and it is,
in the regime $p^2\ll |m(\langle T \rangle)|^2$,
found to yield a result proportional to
\beqa
\Upsilon \sim
\frac{1}{16} \frac{1}{(4 \pi)^2} \; {\cal R} \;
\{ \int^1_0 dx ( \frac{1}{\varepsilon}
- ln \frac{ \Box_0 x(1-x) + |m(\langle T \rangle)|^2}{2 \pi \mu^2} + const )
\}\; {\cal R}
\nonumber\\
\approx \frac{1}{16} \frac{1}{(4 \pi)^2}  ( \frac{1}{\varepsilon}
- ln \frac{ |m(\langle T \rangle)|^2}{2 \pi \mu^2} + const) \; {\cal R}^2
\label{rtwo}
\eeqa
Upon renormalisation, one finds that the term (\ref{rtwo})
is contained in the following superfield expression
\beqa
\Upsilon \sim \int d^4 \theta \bar{R} R \;
ln \frac{{\cal M}(\langle T \rangle)}{\mu} + h.c.
\label{tracemass}
\eeqa
Note that this yields a local contribution proportional to the logarithm of
the holomorphic mass ${\cal M}(\langle T \rangle)$.

\begin{figure}[t]
\[
\begin{picture}(5,60)
 \put(-60,0){\line(10,0){157}}
 \multiput(-105,2)(10,0){5}{\oval(10,10)[b]}
\put(-60,0){\line(3,2){80}}
\put(20,53){\line(3,-2){79}}
\put(160,-2){${\cal R}$}
\put(-130,-2){${\cal R}$}
\multiput(103,2)(10,0){5}{\oval(10,10)[b]}
\multiput(20,52)(0,5){6}{\circle*{2}}
\put(30,70){$K_0$}
\put(28,50){$\bar{\cal F}$}
\put(5,50){${\cal F}$}
\put(-60,10){$A$}
\put(-55,-15){$\bar{A}$}
\put(80,-15){$A$}
\put(85,10){$\bar{A}$}

\end{picture}
\]
\caption{Local non-holomorphic massive contribution to ${\cal R}^2 K_0$}
\label{f5}
\end{figure}
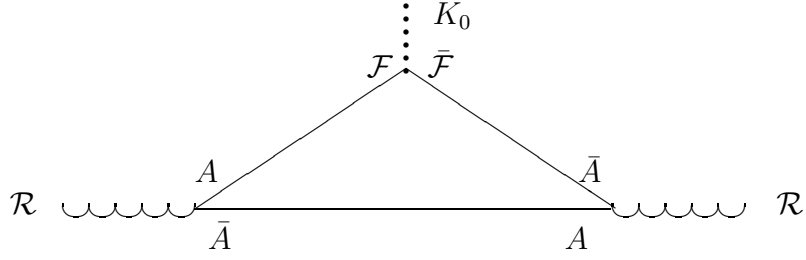

Next, consider the triangle graph depicted in Figure \ref{f5}.
This component graph is contained in the supergraph shown in Figure
\ref{f4}.  The triangle graph in Figure \ref{f5}
has a top vertex given by the coupling
$K_0 {\cal F} \bar{\cal F} $ and bottom vertices given by the
coupling $\bar{A} A {\cal R}$ in (5.3).
Making use of the component propagators \cite{90}
\beqa
<0|T\{ A(x) \bar{A}(x') \}|0> &=& i \Delta_F (x-x') \nonumber\\
<0|T\{ A(x) F(x') \}|0> &=& - i m \Delta_F (x-x') \nonumber\\
<0|T\{ \bar{A}(x) \bar{F(x')} \}|0> &=& -i \bar{m}
 \Box_0 \Delta_F (x-x') \nonumber\\
\Delta_F (x-x') &=& \frac{1}{\Box_0 - |m(\langle T \rangle)|^2}
\eeqa
and reverting to momentum space yields the following expression for the
component graph
\beqa
{\cal R}(q_1)\; {\cal R}(q_2)\; K_0 (q_1+q_2) \;
|m(\langle T \rangle)|^2 \;
\int^0_1 dx
\; \int^{1-x}_0 dy
 \int \frac{dl^{2 \omega}}{(2 \pi)^{2 \omega}} \;\;
\frac{1}
{(l^2 + 2 l\cdot \hat{p} + \hat{M}^2)^3}
\label{int1}
\eeqa
where $\omega = 2- \varepsilon$ and where $q_1$ and $q_2$ denote the two
incoming momenta and $q_1+q_2$ the outgoing momentum.  Also, $\hat{p}=
q_1 x + ( q_1 + q_2) y$ and
$\hat{M}^2 = q_1^2 x + (q_1 + q_2)^2 y + |m(\langle T \rangle)|^2$.
Integral (\ref{int1})
can be evaluated using standard techniques and yields
\beqa
{\cal R}(q_1)\; {\cal R}(q_2)\; K_0 (q_1 + q_2) \;
|m(\langle T \rangle)|^2 \;
\int^0_1 dx \int^{1-x}_0 dy
\frac{\Gamma(3-\omega)}{(\hat{M}^2 - \hat{p}^2)^{3-\omega}}
\eeqa
Working in the regime where $q^2 \ll
|m \langle T \rangle|^2$ yields
\beqa
&&
{\cal R}(q_1)\; {\cal R}(q_2)\; K_0(q_1+q_2) \;
\frac{|m(\langle T \rangle)|^2}{|m(\langle T \rangle)|^2}
\left( \int^0_1 dx \int^{1-x}_0 dy \right)
= \frac{1}{2}
{\cal R}(q_1)\; {\cal R}(q_2)\; K_0 (q_1+q_2) \nonumber\\
&&\rightarrow \frac{1}{2} {\cal R}^2(x)\; K_0(x)
\label{int2}
\eeqa
Thus, the massive component graph \ref{f5} gives rise to a finite
non-holomorphic contribution proportional to ${\cal R}^2(x) K_0(x)$.
Similarly, replacing the ${\cal F} \bar{\cal F}$-coupling to
$\kappa^2 K_0$ by the  ${\cal F} \bar{\cal F}$-coupling to
$- 3 ln {\cal Z}$ yields a local non-holomorphic
contribution proportional to ${\cal R}^2 ln {\cal Z}$.  Both these terms
are contained in the component expansion of
\beqa
{\cal L} &=&  \beta_H \int d^4 \theta \bar{R} R \left( \kappa^2
K_0-3ln{\cal Z} \right)
\label{masschiral}
\eeqa
Note that the component expansion of (\ref{masschiral})
doesn't contain any $\tilde{\cal R}
{\cal R}$-terms.

The massive contributions given above, (\ref{tracemass}) and
(\ref{masschiral}), are written in conventional superspace.
These contributions have to be transformed over into K\"{a}hler superspace
using (\ref{rtrans}). To lowest order in the background fields,
the massive contributions we have singled out in the above discussion
are in K\"{a}hler superspace
again given by (\ref{tracemass}) and (\ref{masschiral}).

Let us now turn to string theory.  Even though the simple field theory
toy-model discussed above doesn't do justice to the complexity of
string theory, it is nevertheless useful in that it points out two
possible types of moduli dependent 1-loop contributions due to the
massive modes, namely local holomorphic contributions proportional
to the logarithm of chiral masses of massive string states as well as
local non-holomorphic contributions proportional to the K\"{a}hler potential
$K$ and to the K\"{a}hler metric ${\cal Z}$.  Thus, it seems reasonable
to parametrise
the local
1-loop contributions due to the massive modes in string theory
in the following way
\beqa
{\cal L}^{massive}&=& \frac{1}{6 \cdot (4 \pi)^2} \sum_I \{ \sum_i \left[
\int d^2 \Theta\;
ln {\cal M}_i(T^I)  \left( \rho_H^{iI} W^2_{\alpha \beta \gamma}
+ \sigma_H^{iI} (\bar{\cal D}^2 - 8 R)
(G^2_{\alpha \dot{\alpha}} - 4 \bar{R} R) \right. \right. \nonumber\\
&+& \left.  \left. \tau_H^{iI}
(\bar{\cal D}^2 - 8 R) {\bar R} R \right) \right] \nonumber\\
&+& \frac{1}{3 \cdot (4 \pi)^2}
\alpha_H^I \int d^4 \theta \left(G^2_{\alpha \dot{\alpha}} -
4 \bar{R} R \right)\, ln(T^I + \bar{T}^I)  \nonumber\\
&+& \frac{1}{3 \cdot (4 \pi)^2}
\beta_H^I \int d^4 \theta \bar{R} R \, ln(T^I + \bar{T}^I)
+ h.c.  \}
\label{masstr}
\eeqa
Note that these are the only local terms quadratic in gravity and
of massive dimension four which do not vanish
in the limit of constant moduli fields $T^I$.  Also note that the
non-holomorphic contributions in (\ref{masstr}) can arise both as
$\bar{R} R$ and $G^2_{\alpha \dot{\alpha}}$-terms and that their
component expansion does not contain
any $\tilde{\cal R}
{\cal R}$-terms.
${\cal L}^{massive}$ is of the type (\ref{loopcoup}) discussed earlier.
The sum in (\ref{masstr}) over massive string states is to be thought
of us being over a restricted set of massive states.  For instance, it may
be thought of as being restricted to untwisted massive states carrying
internal winding and momentum numbers, with dimensionless
chiral masses given by
${\cal M}(T^I) = n + i m T^I$ \cite{FerKouLuZw}.  Then, the
sum runs over all integer
quantum numbers $(n,m)\not= (0,0)$.
Again, let us point out that massive states can contribute local
CP even non-holomorphic terms to the gravitational couplings.

Next, consider the total 1-loop moduli dependent contributions
from both massless and massive states given by the sum of
(\ref{lagmless}) and (\ref{masstr}) and impose invariance under modular
transformations, also refered to as duality invariance.  Care needs
to be taken when specifying the modular group.  In the case where the
underlying $T^6$-torus factorises into a product of three $T^2$'s, the
modular group is given by $(SL(2,Z))^3$.  In most cases, however,
the underlying $T^6$ will not factorise in this way, and the duality
group will be a subgroup of $(SL(2,Z))^3$ not known in most cases.
An example of the latter is the $Z_7$-orbifold \cite{ErlSpa}.

Now, the explicit calculation \cite{Dix,12,13,May,Bailin1}
of the moduli dependent
threshold corrections to the gauge couplings and to the gravitational
coupling proportional to $W^2_{\alpha \beta \gamma}$ shows that only
$N=2$ orbifold sectors, associated with one untwisted complex
$T^2$-plane, contribute to these threshold corrections.  If the
underlying $T^6$ torus factorises into $T^6= T^2  \oplus T^4$ with
the unrotated plane lying in $T^2$, then the threshold corrections are
invariant under $\Gamma = SL(2,Z)$.  If, however, $T^6 \not= T^2  \oplus
T^4$, then the threshold corrections will only be invariant under a subgroup
of $\Gamma$.  Such a subgroup has, in general, a larger set of automorphic
functions, and so the threshold corrections will be composed of different
terms \cite{May,Bailin1,Bailin2}.
Consequently, the exact expression for the threshold
corrections cannot be inferred anymore from (\ref{lagmless}) alone
by demanding duality invariance.  Thus, we will in the following stick
to those orbifolds for which $T^6= T^2 \oplus T^4$ with the untwisted
plane lying in $T^2$.

Let $T$ denote the modulus associated with the untwisted plane.  Then,
invariance of the threshold
corrections under $\Gamma=SL(2,Z)$ requires that
\beqa
\sum_{{\cal M}_i \neq 0} \rho_H^i \, ln {\cal M}_i(T) &=&  \rho_H
\, ln W(T) \nonumber\\
\sum_{{\cal M}_i \neq 0} \sigma_H^i \,
ln {\cal M}_i(T) &=&  \sigma_H
\, ln W(T) \nonumber\\
\sum_{{\cal M}_i \neq 0} \tau_H^i \,
ln {\cal M}_i(T) &=& \tau_H
\, ln W(T) \eeqa
where $W(T)$ denotes an automorphic function of modular weight $-1$
given by \cite{CveLu}
\beqa
W(T)= \frac{H(T)}{{\eta}^2(T)} \eeqa
Here, $H(T)= H(j(T))$ denotes a rational function of
the absolute modular invariant function $j(T)$.  $H(T)$ has the properties
that it is regular inside the fundamental domain of the $T$-plane and
that it possibly
vanishes at some points inside the fundamental domain, e.g.
$H(T) \propto
(j(T) - 1728)$ vanishes at $T=1$.  The zeros of $H$ indicate
that at these points some fields, which contribute to the thresholds,
become massless.
Thus, the
sum of all massless and massive contributions
can be rewritten as
\beqa
{\cal L}^{total} &=& \frac{b}{24 \cdot (4 \pi)^2} \int d^4 \theta
W^2_{\a\b\gamma}
\frac{D^2}{\Box_0} \;\;  ln((T + \bar{T})|W(T)|^{-2}) \nonumber \\
&+&  \frac{p}{24 \cdot (4 \pi)^2} \int d^4 \theta
\left( \bar{\D}^2 - 8 R\right) \left(
G^2_{\a\dota} - 4 \bar{R} R \right)
\frac{D^2}{\Box_0} \;\;
   ln((T  + \bar{T})|W(T)|^{-2})
\nonumber \\
&+&  \frac{h}{24 \cdot (4 \pi)^2} \int d^4 \theta
\left( \bar{\D}^2 - 8 R\right) \left(
\bar{R} R \right)
\frac{D^2}{\Box_0} \;\;
  ln((T + \bar{T})|W(T)|^{-2})
\nonumber \\
&+&  \frac{\alpha_H}{3 \cdot (4 \pi)^2}
 \int d^4 \theta (G^2_{\alpha \dot{\alpha}} - 4 \bar{R} R)\;\;
ln((T + \bar{T})|W(T)|^{-2})
\nonumber \\
&+&   \frac{\beta_H}{3 \cdot (4 \pi)^2} \int d^4 \theta \bar{R} R
\;\; ln((T + \bar{T})|W(T)|^{-2}) + h.c.
\label{total}
\eeqa 
where the requirement of duality invariance imposes the following
relations
\beqa
\rho_H &=& b \nonumber\\
\sigma_H &=& p + \alpha_H \nonumber\\
\tau_H &=& h + \beta_H
\label{heavylight}
\eeqa
Note that (\ref{heavylight})
determines one of the heavy coefficients, namely $\rho_H$,
in terms of the light coefficient $b$ given in (\ref{ncoef}).

Let us next choose a vev for the modulus $T$, $T=\langle T \rangle$.  Then,
(\ref{total}) turns into the local expression
\beqa
{\cal L}^{vev} &=& - \frac{b}{12 \cdot (4 \pi)^2}  \;
ln(\langle T \rangle + \bar{\langle T \rangle})\; \int d^2 \Theta
W^2_{\a\b\gamma} \nonumber\\
&+& \frac{b}{6 \cdot (4 \pi)^2}  \;
lnW(\langle T \rangle)\; \int d^2 \Theta
W^2_{\a\b\gamma} \nonumber\\
&+&  \frac{p + \alpha_H}{3 \cdot (4 \pi)^2} \;
ln((\langle T \rangle + \bar{\langle T \rangle})|W(\langle T \rangle)|^{-2})
\int d^4 \theta (G^2_{\alpha \dot{\alpha}} - 4 \bar{R} R)
\nonumber \\
&+&  \frac{h+ \beta_H}{3 \cdot (4 \pi)^2}  \;
 ln((\langle T \rangle + \bar{\langle T \rangle})
|W(\langle T \rangle)|^{-2}) \int d^4 \theta \bar{R} R
+ h.c.
\label{lvev}
\eeqa
In components, (\ref{lvev}) reads
\beqa
&& {\cal L} =  -\frac{1}{96 \cdot (4 \pi)^2}
 b\; C^{mnpq} C_{mnpq} ln((\langle T \rangle +
\bar{\langle T \rangle})|W(\langle T \rangle)|^{-2})
\nonumber\\
&& + \frac{1}{6 \cdot (4 \pi)^2}
(p + \alpha_H) ({\cal R}^{mn} {\cal R}_{mn} - \frac{1}{3}
{\cal R}^2) ln((\langle T \rangle +
\bar{\langle T \rangle})|W(\langle T \rangle)|^{-2}) \nonumber \\
&&+ \frac{1}{3 \cdot 72 \cdot (4 \pi)^2}
( h+ \beta_H) {\cal R}^2
ln((\langle T \rangle + \bar{\langle T \rangle})
|W(\langle T \rangle)|^{-2}) +\cdots \eeqa
Rewriting this into Gauss-Bonnet and $C^2$-terms yields
\beqa
&& {\cal L} =  -\frac{1}{96 \cdot (4 \pi)^2} (b-8 p -8\alpha_H)
 C^{mnpq} C_{mnpq} ln((\langle T \rangle +
\bar{\langle T \rangle})|W(\langle T \rangle)|^{-2})
\nonumber\\
&&  	-\frac{1}{12 \cdot (4 \pi)^2}
(p + \alpha_H) GB ln((\langle T \rangle +
\bar{\langle T \rangle})|W(\langle T \rangle)|^{-2}) \nonumber \\
&&+ \frac{1}{3 \cdot 72 \cdot (4 \pi)^2}
( h+ \beta_H) {\cal R}^2 ln((\langle T \rangle +
\bar{\langle T \rangle})|W(\langle T \rangle)|^{-2}) +\cdots
\label{vevgb}
\eeqa
Hence, massless and massive contributions are of the same form for constant
vev $\langle T \rangle$.

Recall that, as discussed in section 3, the conventional choice for the
orbifold tree-level Lagrangian (\ref{choice}) translates into the absence of
tree-level couplings proportional to naked $C^2$-terms.
(\ref{vevgb}), on the other hand, shows that for constant moduli background
fields one finds in general naked $C^2$-terms at the one-loop level.  As
argued in the introduction, there is a priori nothing wrong with naked
$C^2$-terms in the effective orbifold Lagrangian.  On the other
hand, it might turn out that
in an actual string calculation the coefficient
$b-8p-8 \alpha_H$ of the naked $C^2$-term is really found to be zero.
 Then
\beqa
8 \alpha_H = b - 8 p
\label{alphaH}
\eeqa
and, hence,
\beqa
\sigma_H = \frac{b}{8}
\eeqa
Then, both $\alpha_H$ and $\sigma_H$ are determined by light coefficients.
Note that, in order
to have absence of naked $C^2$-terms for constant moduli background
fields, the non-holomorphic contributions
due to the massive modes are crucial.  Also note that the coefficient
$(h+ \beta_H)$ remains undetermined.
On the other hand, inserting the coefficient
(\ref{alphaH}) into (\ref{total}) yields the following 1-loop
moduli dependent
corrections to the gravitational couplings associated with the
untwisted $T^2$-plane
\beqa
{\cal L}^{total} &=&
\frac{b-8p}{24 \cdot (4 \pi)^2} \int d^4 \theta
W^2_{\a\b\gamma}
\frac{D^2}{\Box_0}  ln((T + \bar{T})|W(T)|^{-2}) \nonumber \\
&+&\frac{p}{24 \cdot (4 \pi)^2} \int d^4 \theta
\{8 W^2_{\a\b\gamma} +
\left( \bar{\D}^2 - 8 R\right) \left(
G^2_{\a\dota} - 4 \bar{R} R \right) \}
\frac{D^2}{\Box_0} \;\;    ln((T  + \bar{T})|W(T)|^{-2})
\nonumber \\
&+&\frac{h}{24 \cdot (4 \pi)^2} \int d^4 \theta
\left( \bar{\D}^2 - 8 R\right) \left(
\bar{R} R \right)
\frac{D^2}{\Box_0} \;\;  ln((T + \bar{T})|W(T)|^{-2})
\nonumber \\
&+& \frac{ \frac{b}{8} - p}{3 \cdot (4 \pi)^2}
\int d^4 \theta (G^2_{\alpha \dot{\alpha}} - 4 \bar{R} R)\;\;
ln((T + \bar{T})|W(T)|^{-2})
\nonumber \\
&+& \frac{\beta_H}{3 \cdot (4 \pi)^2} \int d^4 \theta \bar{R} R
\;\; ln((T + \bar{T})|W(T)|^{-2}) + h.c.
\label{totalzwie}
\eeqa
where, to repeat, $b$ and $p$ are light coefficients given in
(\ref{ncoef}), $h$ is a light coefficient given in
(\ref{hcoeff}), and where $\beta_H$ denotes one of the heavy
coefficients introduced in (\ref{masstr}).
It would, indeed,
be of importance to compute these coefficients directly
in string theory.
Thus, (\ref{totalzwie}) shows that,
even if there is exact cancellation (\ref{alphaH})
of naked $C^2$ terms for constant moduli backgrounds,
such a cancellation will not hold anymore for arbitrary moduli
backgrounds,
and the effective Lagrangian
will then contain $C^2$-terms with
both non-local and local moduli dependent
functions.

\section{Conclusion}

\setcounter{equation}{0}

\hspace*{.3in}
We reviewed the
manifestly supersymmetric procedure introduced in \cite{15}
for calculating mixed gravitational-
K\"{a}hler and mixed gravitational-$\sigma$ model anomalies
in field theory.
We applied it to
$Z_N$-orbifolds in order to calculate
the 1-loop moduli dependent contributions
to gravitational couplings due to the massless modes
running in the loop.  We introduced a suitable parametrisation of
those massless 1-loop contributions which we didn't compute explicitly,
such as the one from the supergravity multiplet.
Sticking to the conventional choice \cite{GROSS,TSEYT} for the tree-level
gravitational coupling of the dilaton, we then showed in Appendix B that
the massless modes will, in general, contribute
to a naked $C^2$-term, that is to a $C^2$-term which is not contained
in the Gauss-Bonnet combination.  We then also introduced a suitable
parametrisation of the moduli dependent 1-loop contributions to the
gravitational couplings due to the massive modes.
Such contributions may occur in two types, namely as local holomorphic
contributions proportional to the chiral masses of (some restricted set
of) massive states and also as local non-harmonic contributions
to the gravitational couplings of
${\cal R}^2$ and ${\cal R}^{mn} {\cal R}_{mn}$.
Rewriting ${\cal R}^{mn} {\cal R}_{mn}$ into $C^2$ and $GB$-terms shows
that massive modes may contribute local non-harmonic terms to naked
$C^2$-terms.  Imposing duality invariance of the threshold corrections
allows one to relate heavy and light coefficients appearing in the
parametrisation of the massive and the massless contributions, respectively.
To repeat, we see no dynamical reason why the naked $C^2$-terms due to
massless and massive modes should cancel
for constant moduli background fields.  However, we have for
completeness explored the conditions for such a cancellation.
We have shown that,
in the case of constant background fields,
the
contributions to $C^2$ coming from the massive modes can cancel against
the massless contributions.  However, in the case of non-constant moduli
such a cancellation doesn't hold anymore, and the effective Lagrangian
will then contain both non-local and local terms proportional to
$C^2$.  Such terms are interesting in that they might have cosmological
implications.

\section{Acknowledgement}

\hspace*{.3in}  We would like to thank Ignatios Antoniadis,
Jens Erler, Vadim
Kaplunovsky and Jan Louis for fruitful discussions.

\section{Appendix A}

\setcounter{equation}{0}

\hspace*{.3in}
We will now show for the case of the $Z_4$-orbifold
that the field theoretical
calculation of the coefficient $b^I$ (\ref{ncoef}) agrees with
the string scattering amplitude calculation of \cite{13}.
The $Z_4$ gauge group is given by
$G = E_6
\x SU(2) \x U(1) \x E_8$.  The massless spectrum thus
contains 330 vector multiplets.  In addition,
it contains in the untwisted sector 6 moduli fields
with modular weights ${\bf q}$ given by
\beqa
T_{ij} \;\;\; &&{\bf q}=(-2,0,0),(0,-2,0),(-1,-1,0),(-1,-1,0) \;\;\;i,j=1,2
\nonumber\\
T_3 \;\;\; &&{\bf q}=(0,0,-2) \nonumber\\
U_3 \;\;\; &&{\bf q}=(0,0,0)
\label{modu}
\eeqa
as well as matter fields with modular weights ${\bf q}$ given by
\beqa
({\bf 27},{\bf 2}) + ({\bf 1},{\bf 2}) \;\;\; &&{\bf q}=(-1,0,0)\nonumber\\
({\bf 27},{\bf 2}) + ({\bf 1},{\bf 2}) \;\;\; &&{\bf q}=(0,-1,0)\nonumber\\
({\bf 27},{\bf 2}) + ({\bar{\bf 27}},{\bf 1}) \;\;\; &&{\bf q}=(0,0,-1)
\eeqa
The twisted sector $\Theta=(\frac{1}{4}, \frac{1}{4},\frac{1}{2})$
contains the following matter multiplets
\beqa
16 \;\; ({\bf 27},{\bf 1}) \;\;\; &&{\bf q}=(- \frac{3}{4}, -\frac{3}{4},
- \frac{1}{2})\nonumber\\
16 \;\; ({\bf 1},{\bf 2}) \;\;\; &&{\bf q}=(- \frac{7}{4},- \frac{3}{4},
-\frac{1}{2})\nonumber\\
16 \;\; ({\bf 1},{\bf 2}) \;\;\; &&{\bf q}=(-\frac{3}{4}, -\frac{7}{4},
-\frac{1}{2})\nonumber\\
16 \;\; ({\bf 1},{\bf 1}) \;\;\; &&{\bf q}=(-\frac{11}{4}, -\frac{3}{4},
-\frac{1}{2})\nonumber\\
16 \;\; ({\bf 1},{\bf 1}) \;\;\; &&{\bf q}=(-\frac{7}{4}, -\frac{7}{4},
-\frac{1}{2})\nonumber\\
16 \;\; ({\bf 1},{\bf 1}) \;\;\; &&{\bf q}=(-\frac{3}{4}, -\frac{11}{4},
-\frac{1}{2})\nonumber\\
16 \;\; ({\bf 1},{\bf 1}) \;\;\; &&{\bf q}=(-\frac{3}{4}, -\frac{3}{4},
-\frac{3}{2})\nonumber\\
16 \;\; ({\bf 1},{\bf 1}) \;\;\; &&{\bf q}=(-\frac{3}{4}, -\frac{3}{4},
\frac{1}{2})
\eeqa
The twisted sector ${\Theta}^2=(\frac{1}{2}, \frac{1}{2},0)$
contains the following matter multiplets
\beqa
10 \;\; ({\bf 27},{\bf 1}) \;\;\; &&{\bf q}=(-\frac{1}{2}, -\frac{1}{2}, 0)
\nonumber\\
6 \;\; (\bar{{\bf 27}},{\bf 1}) \;\;\;
&&{\bf q}=(-\frac{1}{2}, -\frac{1}{2}, 0)
\nonumber\\
10 \;\; ({\bf 1},{\bf 2}) \;\;\; &&{\bf q}=(-\frac{3}{2}, -\frac{1}{2}, 0)
\nonumber\\
10 \;\; ({\bf 1},{\bf 2}) \;\;\; &&{\bf q}=(-\frac{1}{2}, -\frac{3}{2}, 0)
\nonumber\\
6 \;\; ({\bf 1},{\bf 2}) \;\;\; &&{\bf q}=(-\frac{1}{2}, \frac{1}{2},0)
\nonumber\\
6 \;\; ({\bf 1},{\bf 2}) \;\;\;
&&{\bf q}=(\frac{1}{2}, -\frac{1}{2}, 0)\nonumber\\
16 \;\; ({\bf 1},{\bf 1}) \;\;\; &&{\bf q}=(-\frac{1}{2}, -\frac{1}{2},0)
\eeqa
The field theoretical coefficients $b^I$ given in (\ref{ocoef})
are then readily evaluated to be equal to
\beqa
b^1 = b^2 = - 720 = -30 \cdot 24 \;\;,\;\;
b^3 = 264 = 11 \cdot 24
\label{b123}
\eeqa
The Green-Schwarz coefficients $\delta^I_{GS}, I=1,2,$ for the $Z_4$-orbifold
are known from the study of the running of gauge couplings and are
given by \cite{DFKZ}
\beqa \delta^1_{GS} = \delta^2_{GS} = - 30 \eeqa
Taking them into account yields the following
modified coefficients $b^I$ (\ref{ncoef})
\beqa
b^1 = b^2 = 0 \;\;,\;\; b^3 = 11 \cdot 24
\label{bfield}\eeqa

The string calculation \cite{13} of the coefficients
$b^I$, on the other hand, shows that only in the case where there are
orbifold
sectors with one untwisted complex plane is the associated coefficient
$b$ non-vanishing.  Since these sectors possess two spacetime supersymmetries,
they are called $N=2$-sectors.  The $Z_4$-orbifold possesses one such $N=2$
sector, namely the ${\Theta}^2$-twisted sector.  $\Theta^2=(\frac{1}{2},
\frac{1}{2},0)$ leaves the third complex plane untwisted, and thus only
$b^3$ will be non-vanishing.  For each $N=2$ sector there is an associated
$(2,2)$ symmetric $Z_M$-orbifold with $N=2$ spacetime supersymmetry.  In
the case of the $Z_4$-orbifold under consideration, the associated
$N=2$ orbifold is a $Z_2$-orbifold obtained by twisting the underlying
$T^6$-torus with $\Omega = \Theta^2 = (\frac{1}{2}, \frac{1}{2}, 0)$.
This $N=2$ $Z_2$-orbifold has gauge group
$\tilde{G}=E_7 \x SU(2) \x E_8$.  Its massless spectrum
is given by the following $N=2$ multiplets \cite{FerPor,Jens}.
The untwisted moduli multiplets are given by
2 vector multiplets and
4 hyper multiplets.  The untwisted matter
multiplets are given by
$({\bf 56},{\bf 2})$ hyper multiplets.
Since $dim \tilde{G} = 384$, there are in addition 384 vector
multiplets.

The twisted sector $\Omega = (\frac{1}{2}, \frac{1}{2}, 0)$ contains
8 $({\bf 56}, {\bf 1})$ hyper multiplets and 32 $({\bf 1},{\bf 2})$
hyper multiplets \cite{Jens}.  Thus, there are a total
of 386 $N=2$ vector multiplets and of 628 $N=2$ hyper multiplets.
The massless spectrum contains, in addition, one $N=2$ supergravity multiplet
as well as one $N=2$ dilaton multiplet.  Each of these $N=2$ multiplets
contributes an amount $(3 \beta - {\beta}')$ given in Table
\ref{anomtab} to the
on-shell trace anomaly $T^m_m = -\frac{1}{24} \frac{1}{(4 \pi)^2}
(3\beta - {\beta}')C^2$.  Then, the string calculation \cite{13}
yields the non-vanishing $b^3$-coefficient as a sum over the trace anomaly
contribution of all the massless $N=2$ multiplets of the $Z_2$-orbifold
\beqa
b^3 =  \sum_s (3 \beta_s -\beta'_s) \eeqa
It is computed to be
\beqa
b^3 = \frac{2}{15} [ \frac{15}{2} 628 - \frac{15}{2} 386
+ \frac{165}{2} + \frac{330}{4} ] = 11 \cdot 24 \eeqa
which agrees with the result (\ref{bfield}) of the field theory
calculation.

\begin{table}
\[
\begin{array}{|c|c|c|c|} \hline
\mbox{Field}  & \mbox{ $\frac{15}{2}(3
\beta - \beta')$} &\mbox{N=2 multiplet}
& \mbox{ $\frac{15}{2} (3 \beta - \beta')$}
\\ \hline \hline
real \;\; scalar & 1 &  hyper & \frac{15}{2} \\ \hline
Weyl \;\; fermion & \frac{7}{4} & vector & -\frac{15}{2} \\ \hline
vector \;\; field & -13 & sugra & \frac{165}{2} \\ \hline
graviton & 212 & dilaton & - \frac{330}{4} \\ \hline
gravitino & - \frac{233}{4} &  & \\ \hline
antisymmetric \;\; tensor & 91 &  &  \\ \hline
\end{array}
\]
\caption{Trace anomaly coefficient $(3 \beta - \beta')$ for various
component fields and $N=2$ multiplets in 4D [34,37,12].}
\label{anomtab}
\end{table}

\section{Appendix B}

\setcounter{equation}{0}

\hspace*{.3in}
We will, in this appendix, show
that the coefficients $b^I-8p^I$ of the naked $C^2$-term cannot be set
to zero in
the $Z_4$ orbifold by an appropriate
choice of the unknown coefficient $\xi$.  Thus, assuming that
there is no Green-Schwarz removal other than of a Gauss-Bonnet combination,
it appears that the massless sector of an orbifold theory will in general
contribute naked $C^2$-terms to gravitational couplings.

The coefficients $b^I - 8p^I$ were found to be given by
\beqa
b^I - 8 p^I
&=& 21 + 1 + n_M^I - {\rm dim}G + \sum_i (1+2q^i_I)  \nonumber\\
&-& \{- 3 dim \; G - 1
- \frac{1}{3} (\sum_i 1) + 8 \xi \}
\label{bipi}
\eeqa
Setting these coefficients to zero yields
\beqa
8 \xi &=& 21 + 1 + n_M^I +2 {\rm dim}G + \sum_i (1+2q^i_I)
+ 1 + \frac{1}{3} (\sum_i 1)
\label{coea}
\eeqa
where, again, $n_M^I = 3+ 2 q^I = -1$.
Note that the left-hand side of this expression
 should be plane independent.
The right-hand side, on the other hand,
contains a piece which looks plane dependent and is
given by
\beqa
8 \tilde{\xi} &=& 2 \sum_i q^i_I
\label{tilchi}
\eeqa
One should thus check whether
the two inequivalent planes
(planes 2 and 3) of the $Z_4$-orbifold give rise to the same
amount (\ref{tilchi}).  The plane dependent terms (\ref{tilchi})
are, however, entirely coming from the $b^I$-coefficient in (\ref{bipi}).
As shown in (\ref{b123}) of
Appendix A, these $b^I$-coefficients are indeed
plane dependent and, hence, planes 2 and 3 give rise to different
amounts (\ref{tilchi}).
  Thus, the coefficients
$b^I - 8 p^I$ cannot be simultaneouly set to zero by an appropriate
choice of $\xi$.

\end{document}